\documentclass[aps, prc, twocolumn, groupedaddress]{revtex4}
\usepackage{graphicx}
\usepackage{tabularx}
\newcolumntype{Y}{>{\centering\arraybackslash}X}
\pdfoutput=1 
\usepackage{amsmath}
\usepackage{natbib}

\usepackage{color}
\usepackage{subfigure}
\usepackage{url}
\usepackage{verbatim}
\usepackage{lipsum}
\usepackage{braket}

\begin{document}

\title{Quantum Simulations of Nuclei and Nuclear Pasta with the Multi-resolution Adaptive Numerical Environment for Scientific Simulations}
\author{I. Sagert$^1$, G. I. Fann$^2$, F. J. Fattoyev$^1$, S. Postnikov$^1$, C. J. Horowitz$^1$}
\affiliation{$^1$ Center for Exploration of Energy and Matter, Indiana University, Bloomington, Indiana, 47308, USA\\
$^2$Computer Science and mathematics Division, Oak Ridge National Laboratory, Oak Ridge, Tennessee 37831, USA}
\date{\today}

\begin{abstract}
\textbf{Background:} Neutron star and supernova matter at densities just below the nuclear matter saturation density is expected to form a lattice of exotic shapes. These so-called nuclear pasta phases are caused by Coulomb frustration. Their elastic and transport properties are believed to play an important role for thermal and magnetic field evolution, rotation and oscillation of neutron stars. Furthermore, they can impact neutrino opacities in core-collapse supernovae. \textbf{Purpose:} In this work, we present proof-of-principle 3D Skyrme Hartree-Fock (SHF) simulations of nuclear pasta with the Multi-resolution ADaptive Numerical Environment for Scientific Simulations (MADNESS). \textbf{Methods:} We perform benchmark studies of $^{16} \mathrm{O}$, $^{208} \mathrm{Pb}$ and $^{238} \mathrm{U}$ nuclear ground states and calculate binding energies via 3D SHF simulations. Results are compared with experimentally measured binding energies as well as with theoretically predicted values from an established SHF code. The nuclear pasta simulation is initialized in the so-called \textit{waffle} geometry as obtained by the Indiana University Molecular Dynamics (IUMD) code. The size of the unit cell is 24\:fm with an average density of about $\rho = 0.05 \:\mathrm{fm}^{-3}$, proton fraction of $Y_p = 0.3$ and temperature of $T=0$\:MeV. \textbf{Results:} Our calculations reproduce the binding energies and shapes of light and heavy nuclei with different geometries. For the pasta simulation, we find that the final geometry is very similar to the initial \textit{waffle} state. We compare calculations with and without spin-orbit forces. We find that while subtle differences are present, the pasta phase remains in the waffle geometry. \textbf{Conclusions:} Within the MADNESS framework, we can successfully perform calculations of inhomogeneous nuclear matter. By using pasta configurations from IUMD it is possible to explore different geometries and test the impact of self-consistent calculations on the latter.  
\end{abstract}

\maketitle
\section{Introduction}
\label{intro}
In the high-density environment of neutron star interiors and core-collapse supernovae, nuclear matter is expected to assume a variety of exotic shapes at the liquid-gas phase transition. The different configurations are created by an interplay between the repulsive long-range Coulomb force and a short-range attractive nuclear force \cite{Baym71, Ravenhall83, Hashimoto84}. At densities of around $\rho \sim 0.01 \: \mathrm{fm}^{-3}$, spherical clusters of nuclear matter form a lattice surrounded by electron gas and neutron liquid. In a simple picture, with increasing density, the spheres merge into tubes that eventually transform into plates. As the density increases further, nuclear matter and neutron matter switch their roles resulting in tubes of neutron liquid and, at higher densities, bubbles enclosed by nuclear matter. For $\rho \gtrsim 0.12  \: \mathrm{fm}^{-3}$, the neutron star interior is composed of a homogeneous mixture of neutrons, protons and electrons. In addition to the above shape sequence, many non-trivial geometries can be present and their similarity with different types of pasta (e.g. spheres = gnocci, tubes = spaghetti, planes = lasagna) lead to the terminology \textit{nuclear pasta phases}. \\
The relevant region for nuclear pasta in neutron stars lies between the outer core and the inner crust. Although the radial width of this region is only several hundred meters (in comparison to a neutron star radius of about 10\:km) its thermal and deformational properties can impact neutron star cooling \cite{Lorenz93, Gusakov04}, oscillations \cite{Gearheart11}, spin \cite{Levin01} and magnetic field evolution \cite{Pons13, Horowitz15}. Understanding the physical characteristics of nuclear pasta is therefore an important step towards a correct interpretation of neutron star observables in connection with nuclear matter properties and equation of state. For core-collapse supernovae (CCSN), pasta phases can form in the collapsing stellar iron core and during the post-bounce phase in the proto-neutron star \cite{Sonoda08}. The latter is the hot and compressed stellar core which is formed during the CCSN and left behind after the explosion. Neutrinos that diffuse from the proto-neutron star interior play a crucial role for the CCSN explosion mechanism \cite{Bethe90, Janka12}. The knowledge of the neutrino mean free path in hot nuclear matter --- that can be modified by nuclear pasta \cite{Horowitz04, Horowitz04b} --- is very important in numerical CCSN studies. In addition, pasta phases could have an impact on nucleosynthesis in CCSN and neutron star binary mergers \cite{Caplan14}.\\ 
\newline
Different approaches are taken to study pasta phases. These include calculations in the liquid-drop model \cite{Ravenhall83, Pethick87, Nakazato09}, Thomas-Fermi and Wigner-Seitz cell approximations \cite{Williams85, Lassaut87, Lorenz93, Oyamatsu93, Maruyama05}, molecular dynamics (MD) and quantum molecular dynamics \cite{Horowitz08, Watanabe05, Watanabe09} studies, static Hartree-Fock \cite{Magierski02, Goegelein07, Newton09, Pais12} and time-dependent Hartree-Fock simulations \cite{Schuetrumpf13, Schuetrumpf14}. Studies are usually performed in the so-called unit cell  filled with neutrons, protons and electrons with specific symmetry assumptions and boundary conditions. The pasta matter is then described as an infinite lattice of unit cells. By studying different configurations in the latter and comparing their total energies the ground state can be identified as the configuration with the lowest energy. For numerical studies, it is important to note the non-trivial role of the simulation volume. As only periodic geometries that fit into the unit cell can be explored, the size of the simulation cell must be sufficiently large to at least contain one period of the pasta structure. Even if the latter is fulfilled, effects of the finite volume such as dependence on the simulation space geometry \cite{Molinelli15} and numerical shell effects  \cite{Newton09} can appear. As a consequence, the simulation volume has to be maximized to ensure that finite size effects are minimal. However, this usually comes with a significant increase in computational costs.\\
\newline
The advantage of MD studies lies in their ability to simulate large systems where the length of the simulation space is several hundred fm \cite{Horowitz08, Schneider13, Schneider14} and therefore exceeds the size of a unit cell. This allows to study pasta structures that are less bound to the geometry and boundary conditions of the volume. Furthermore, it is possible to explore bulk properties such as electrical and thermal conductivities, shear and bulk moduli. However, although MD approaches can include quantum effects, the nucleon interaction is typically given by a schematic two-body potential. For self-consistent quantum calculations that account for Pauli blocking, spin-orbit forces and nucleon pairing, Skyrme Hartree-Fock-Bogolyubov (SHF(B)) and density functional theory (DFT) simulations are usually performed. A current drawback of these methods is  their high computational cost. The consequence of the latter is that the system size that can be studied in SHF and DFT calculations is typically much smaller than for MD methods \cite{Pais12, Schuetrumpf13, Pais14, Schuetrumpf14, Schuetrumpf15, Pais15}. Therefore, for large-scale simulations, the applied numerical framework needs to be highly parallelized and should scale well. The Multi-resolution ADaptive Numerical Environment for Scientific Simulations (MADNESS) has been developed to efficiently solve these type of problems exactly and is designed to run on modern supercomputer facilities. With that, our aim is to apply MADNESS to perform large-scale 3D SHF simulation of nuclear pasta. In the current work we introduce our approach and perform benchmark studies of nuclear ground states. We then perform first pasta calculations using MADNESS and a converged MD simulation as starting point. \\
The paper is structured as follows: We first give an overview of the Skyrme Hartree-Fock approach and the MADNESS computational environment in sections \ref{shf} and \ref{madness}, respectively. We continue with a description of how we solve the SHF equations in MADNESS in section \ref{equations} and then present our results for nuclear ground states in section \ref{ground}. In section \ref{pasta} we show our first nuclear pasta simulations with MADNESS and close with a summary in section \ref{summary}. 
\section{Skyrme Hartree-Fock calculations}
\label{shf}
In this section we provide a brief overview of the Skyrme Hartree-Fock method. For a more detailed discussion see e.g. \cite{Maruhn14, Stone07, Vautherin72}. Instead of solving the Schroedinger equation for the A-nucleon wavefunction and the corresponding Hamiltonian $\hat{H}$, the Hartree-Fock approach approximates the ground state of the nuclear configuration by a single Slater determinant $\Psi$. The latter is formed by a complete orthonormal set of single-particle wavefunctions $\psi_i (\vec{r}_i)$ whereas $\vec{r}_i$ contains the spatial, spin, and iso-spin coordinates of the $i$th state. The Slater determinant is obtained via the variational principle by the minimization of the energy expectation value: 
\begin{equation}
\delta E ( \Psi ) = \delta \bra{\Psi} H \ket{\Psi}  = 0.
\end{equation}
As a result, the many-body Schroedinger equation is turned into A single-body Schroedinger equations with the Hartree-Fock mean-field potential $U_\mathrm{HF}$. For each single particle state $\psi_i$, the corresponding HF equation reads:
\begin{align}
H \:  \psi_i (\vec{r}) = \left( -  \frac{\hbar^2}{2m_i} \Delta + U_{\mathrm{HF},q} (\vec{r}) \right) \psi_i (\vec{r}) = E_i \: \psi_i (\vec{r}), 
\label{HF_eq}
\end{align}
where $m_i$ is the nucleon mass and $E_i$ is the energy of the single-particle state. For an iso-spin state $q=n,p$ ($n$=neutrons, $p$=protons), the Hartree-Fock potential $U_{\mathrm{HF},q}$ contains the following contributions: 
\begin{align} 
U_{\mathrm{HF},q} = U_{q,\mathrm{sky}} + U_{q, \mathrm{meff}} + U_{q, so} + U_{q, \mathrm{current}} + U_{q, \mathrm{spin}} .
\label{pot_op}
\end{align}
Furthermore, for protons, the Coulomb potential $U_\mathrm{C}$ and Coulomb exchange potential $U_{C,\mathrm{ex}}$ are added. The first term in eq.(\ref{pot_op}) is the Skyrme potential and can be expressed as
\begin{align}
U_{q,\mathrm{sky}}  &=  b_0 \rho - b^{\prime}_0 \rho_q + b_1 \tau - b^{\prime}_1 \tau_q - b_2 \Delta \rho  \nonumber\\
                       &+ b^{\prime}_2 \Delta \rho_q + b_3 \frac{\alpha+2}{3} \rho^{\alpha + 1} - b^{\prime}_3 \frac{2}{3} \rho^\alpha \rho_q  \nonumber\\
                       &- b^{\prime}_3 \frac{\alpha}{3} \rho^{\alpha - 1} \left( \rho^2_n + \rho^2_p \right) - b_4 \nabla \cdot \vec{J} - b_4^\prime \nabla \cdot \vec{J}_q.              
                       \label{Uskyrme}       
\end{align}
Here, $\alpha$, $b_j$ and $b^{\prime}_j$ ($j=0...4$) are constants specific to the Skyrme potential. They are fitted to reproduce known properties of finite nuclei and infinite nuclear matter. The nucleon number densities $\rho_q$ and kinetic densities $\tau_q$ in eq.(\ref{Uskyrme}) are given by: 
\begin{align}
\rho_q &= \sum_{i}^{N_q} \sum_s | \psi_{i,s} (\vec{r}) |^2,\: \: \tau_q = \sum_{i}^{N_q} \sum_s | \nabla \psi_{i,s} (\vec{r}) |^2, \\
N_p &= Z, \: N_n = A-Z,   \nonumber
\end{align}
where $A$ is the mass number, $Z$ the charge number of the nuclear configuration and $s$ marks the spin of the state. The divergence of the spin-orbit density $\vec{J}$ is determined by:
\begin{align}
\nabla \cdot \vec{J} = -\mathrm{i} \sum_i \sum_{s s^\prime} \left( \nabla \psi_{i,s^\prime}^\star \times \nabla \psi_{i,s}  \cdot  \Braket{ s^\prime  | \vec{\sigma} | s} \right).
\label{divJ}
\end{align}
where $\sigma_i$ are the Pauli matrices. The Coulomb exchange potential $U_{C,\mathrm{ex}}$ in eq.(\ref{pot_op}) can be calculated via the so-called Slater approximation \cite{Chamel08}:
\begin{align}
U_\mathrm{C,ex} (\vec{r})= - e^2 \left( \frac{3}{\pi} \: \rho_p (\vec{r})  \right)^{1/3}.
\end{align}
The Coulomb potential of the protons is given by:
\begin{align}
U_{Cp} =  \int \frac{e^2 \: \rho_p (\vec{s})}{\left| \vec{r} - \vec{s} \right| } \: d \vec{s} . 
\end{align}
Furthermore, when studying nuclear pasta phases, we assume that the unit cell is charge neutral and contains neutrons, protons and electrons. The latter are given via the so-called Jellium approximation and form a background of homogeneous negative density $\rho_J = - Z/V$. As a consequence, in addition to proton Coulomb potential, we have to consider the interaction of the protons with the Coulomb potential of the Jellium:
\begin{equation}
U_{CJ} = \int  \frac{e^2 \: \rho_J}{\left| \vec{r} - \vec{s} \right| } \: d \vec{s} .
\end{equation}
We can sum both contributions to a total Coulomb potential: 
\begin{align}
 U_{Cp} + U_{CJ} &= \int \frac{e^2  \: \left(  \rho_p (\vec{s}) +  \rho_J \right) }{\left| \vec{r} - \vec{s} \right| } \: d\vec{s}, \\
\rightarrow U_{C} & =   \int \frac{e^2 \: \rho_C (\vec{s})}{\left| \vec{r} - \vec{s} \right| }  \: d\vec{s} , 
\end{align}  
where
\begin{align}
U_C = U_{Cp} +  U_{CJ}, \: \rho_C (\vec{r})= \rho_{p} (\vec{r})+ \rho_{J}. 
\label{VC_rhoC}
\end{align}
We will apply the Jellium approximation whenever studying volumes with periodic boundary conditions. The remaining components of the nucleon potential in eq.(\ref{pot_op}) are a contribution that accounts for the effective nucleon mass:
\begin{equation}
U_{q,\mathrm{meff}} =  - \nabla \cdot \left( b_1 \rho - b^{\prime}_1 \rho_q \right) \nabla, 
\label{Vmeff}
\end{equation}
and the spin-orbit potential:
\begin{align} 
U_{q,so} = \mathrm{i} \vec{W}_q \cdot \left( \vec{\sigma} \times \nabla \right) , \:\:\: \vec{W}_q =  \left(b_4 \nabla \rho + b_4^\prime \nabla \rho_q \right)
\label{Vso}
\end{align}
In this work we are focusing on time-independent HF calculations of even-A and even-even nuclei as well as pasta phases. Therefore, we do not include the current and spin operators $U_{q, \mathrm{current}}$ and $U_{q, \mathrm{spin}}$, respectively. These contribute to the Hamiltonian only in case of odd-A and odd-odd nuclei, and when dynamical effects come into play \cite{Bender03}. \\
Since the single particle states $\psi_i$ in eq.(\ref{HF_eq}) depend on the potentials and densities that in turn are derived from the wavefunctions, HF problems have to be solved iteratively with an assumption about the initial single-partilce states $\psi_i$, e.g., harmonic oscillator states, 3D gaussians or plane waves. From these, we derive densities and potentials that are used in the SHF equations to determine a new set of updated states. The calculations are repeated until the solution for the wavefunctions is self-consistent. Due to the large number of states in nuclear pasta simulations, we require a numerical framework that is computationally efficient and parallelized. We therefore apply the Multi-resolution ADaptive Numerical Environment for Scientific Simulations (MADNESS) which will be described in the next section.    
\section{Multi-resolution Adaptive Calculations}
\label{madness}
MADNESS is a numerical framework designed to efficiently solve problems involving integral and partial differential equations in many dimensions. Examples include Hartree-Fock and density functional theory calculations of chemistry and nuclear physics problems \cite{Pei14, Pei12, Fann09, Fann07, Harrison05, Harrison04, Yanai04} with a recent application in studying finite nuclei via solving the HFB equations \cite{Pei14}. Operations in MADNESS are highly parallelized via a combination of MPI and pthreads parallel computing.\\
In MADNESS, functions and operators are described by adaptive pseudo-spectral approximations that are based on a multi-wavelet basis. The latter is given by discontinuous Alpert's multi-wavelets \cite{Alpert93, Alpert02} with Legendre polynomials being applied as scaling functions. Both, scaling functions and multi-wavelets, have disjoint support and are efficient in describing discontinuities and regions with high curvature. Furthermore, with each operator and function having its own adaptive structure of refinement, the user can achieve a defined finite but guaranteed precision. In the following, we will briefly describe the multi-resolution approach in MADNESS whereas details can be found in e.g. \cite{Alpert02, Yanai04, Fann10}. \\
MADNESS projects functions and operators from the user space with a defined width onto a solution interval $[0,1]$. Here, $k$ orthonormal Legendre scaling functions 
\begin{align}
 \phi_i (x) &= \left\{ \begin{array}{l l} 
\sqrt{2i + 1} P_i (2x -1) & \quad \text{for $ 0 \leq x \leq 1$ }\\
0 & \quad \text{otherwise}
\end{array} \right.\\
i &= 0, ..., k-1 \nonumber
\end{align}
can be defined. They are the $i$th Legendre polynomials $P_i (x)$ shifted to $[0,1]$ and normalized. The solution interval is repeatedly cut in half. At level $n$, there are $2^n$ boxes of size $2^{n-1}$. The functions $\phi_i (x)$ are scaled to level $n$ and translated to each subinterval $l$ with size $[2^{-n} l , 2^{-n} (l+1) ]$ where they are given by: 
\begin{align}
\phi_{il}^n (x) &= \sqrt{2^{n}} \phi_i (2^n x - l), \\
i &= 0, ..., k-1, \: l=0,...,n-1 \nonumber
\end{align}
The scaling functions are orthonormal on the interval $[2^{-n} l, 2^{-n} (l+1)]$ and span the sub-spaces $V_n^k$ which form a ladder:  
\begin{align}
V_0^k \subset V_1^k \subset V_2^k \: ... \subset V_n^k  \subset ...
\end{align}
Due to this relation, scaling functions at level $n$ can be derived by scaling functions at level $n+1$ by the two-scale relationship \cite{Jawerth94}. In the \textit{reconstructed} form, a function $f$ that is smooth at level $n$, can be represented by scaling functions $\phi_{il}^n$ and coefficients $s_{il}^n$ as:
\begin{align}
f^n (x) &= \sum_{l=0}^{2^n - 1} \sum_{i=0}^{k-1} s_{il}^n \: \phi_{il}^n (x),  \\
s_{il}^n &= \int_{2^{-n}l}^{2^{-n} (l+1)} \phi_{il}^n (x) \: f(x) dx.
\end{align}
The complementary subspace to $V_n^k$ in $V_{n+1}^k$ is $W_n^k$ with:
\begin{align}
W_n^k \oplus V_n^k = V_{n+1}^k .
\end{align} 
It is spanned by multi-wavelets $\psi_{il}^n (x)$ on the interval $[2^{-n} l, 2^{-n} (l+1)]$ that are obtained by dilation 
and translation of $\psi_i$: 
\begin{align}
\psi_{il}^n (x) &= \sqrt{2^n} \psi_i (2^n x -l), \\
i &= 0, ..., k-1, \:\:\: l=0,...,2^n-1 \nonumber
\end{align}
which, in turn can be derived from the multi-scaling functions by the two-scale relations. Alpert's wavelets are orthonormal within and between scales. Since $V_n^k$ can be decomposed into: 
\begin{align}
V_n^k &= V_0^k \oplus W_0^k \oplus W_1^k \oplus \: ... \oplus  W_{n-1}^k, 
\end{align}
the function $f^n$ can be given as a sum over scaling functions at the coarsest level and wavelets at finer length-scales:
\begin{align}
f^n (x) &= \sum_{i=0}^{k-1} \left( s_{i0}^0 \: \phi_{i} (x) + \sum_{m=0}^{n-1} \sum_{l=0}^{2^m -1} d_{il}^m \: \psi_{il}^m (x) \right), \\
d_{il}^m &= \int_{2^{-m}l}^{2^{-m}(l+1)} f(x) \psi_{il}^m dx
\end{align} 
This is the so-called \textit{compressed} form. The \textit{reconstructed} representation and \textit{compressed} representation are two equivalent forms of $f^n$. For some numerical operations it is better to use the scaling function representation while for others (e.g. inner product of functions) the wavelet form is more efficient. The transformation between both representations of $f^n$ is an orthogonal transformation, it is therefore numerically stable and fast. Going from 1D to 3D, functions are given by tensor products of multi-wavelets and scaling functions are given in the non-standard form. Adaptive refinement is performed locally if the local error is above a truncation threshold $\epsilon$. In 1D, it is accomplished by truncation of small wavelet coefficients whereas MADNESS offers different truncation criteria. For an accurate representation of functions and their derivatives, which we are interested in, coefficients for level $n$ and sub-interval $l$ are neglected when:
\begin{align}
|| d_l^n ||_2 = \sum_i \sqrt{ | d_{il}^n |^2} \leq \epsilon \: \min (1, 2^{-n} L), 
\end{align}
where $L$ is the minimum width of the simulation volume and $\epsilon$ is the desired precision. In MADNESS, Green's functions are represented via low-separation rank expansion in terms of Gaussians. For example the Yukawa kernel is:
\begin{align}
\frac{e^{-k \: r}}{r} = \sum_{m=1}^M \omega_m e^{-p_{1,m} x_1^2} e^{-p_{2,m} x_2^2} e^{-p_{3,m} x_3^2} + O \left( \frac{\epsilon}{r} \right) .
\label{HMkernel}
\end{align}
This reduces a 3D convolution to a set of uncoupled 1D convolutions with $M$ depending on the user-determined precision $\epsilon$. Transformation matrices with respect to the multi-wavelets are pre-computed which allows a fast computation of the convolution. 
\section{Solving the Skyrme Hartree-Fock equations}
\label{equations}
Our general strategy is to rewrite a given differential problem into an integral form and solve it iteratively via convolutions with Green's functions. Correspondingly, we rearrange each HF equation in eq.(\ref{HF_eq}) into their Lippmann-Schwinger form: 
\begin{align}
\left( \alpha^{-1} \Delta + E_i \right) \psi_i (\vec{r}) &= U_\mathrm{HF,q} (r) \: \psi_i (\vec{r}), 
\label{conv1}
\end{align}
where $\alpha^{-1} =  \hbar^2/2m$. This can now be expressed as a convolution with the Green's function for the bound state Helmholtz (BSH) equation:
\begin{align}
\psi_i (\vec{r}) &= - \alpha \:  G_{\mathrm{BSH,i}} \star \left(U_\mathrm{HF,q} \: (\vec{r}) \psi_i (\vec{r}) \right) \\
&= - \alpha \int_{-\infty}^{\infty} G_{\mathrm{BSH,i}} (\vec{r}, \vec{s} ) \left( U_\mathrm{HF,q} \: (\vec{s}) \psi_i (\vec{s}) \right) \: d\vec{s} \\
G_{\mathrm{BSH,i}} & ( \vec{r} , \vec{s} ) = \frac{1}{ 4 \pi |\vec{r} - \vec{s}|} e^{ - k \: | \vec{r} - \vec{s} |} , \: \: k = \sqrt{ - \alpha E_i} .
\end{align}
Similarly, the Coulomb potential $U_C$ for a given total charge density $\rho_C$ and   
\begin{align}
\Delta U_C = - 4 \pi \rho_C 
\label{laplace}
\end{align}
is given by the convolution with the corresponding Green's function $G_C (\vec{r}, \vec{s}) = 1/|\vec{r} - \vec{s}|$. \\
To smooth out possible numerical noise in first and second derivative terms, we apply Gaussian smoothing. Due to its high resolution and adaptive refinement, MADNESS resolves small discontinuities which could then propagate from e.g. the Skyrme potential into the updated wavefunctions. If not damped, the noise can amplify with each iteration. As a consequence, we convolute kinetic densities $\tau_q$ and density laplacians $\Delta \rho_q$ with Gaussians of the following form: 
\begin{align}
f(r) = \left( \sigma \sqrt{2 \pi} \right)^{-3} \: e^{-r^2/(2 \sigma^2)} .
\end{align}
where typically $0.25\:$fm is chosen for $\sigma$. Smoothing is applied for the initial iterations and removed once the configuration starts to converge.\\
\newline 
Our code is based on a previous algorithm to solve the Lippman-Schwinger equations for HF problems with spin-orbit potential \cite{Fann09}. The iterations are therefore performed in a similar fashion. We start out with a set of single particle states at iteration $n=0$ - $\psi_i^n$. These states are ortho-normalized using the a LAPACK hermitian eigensolver for the generalized eigen-system problem \cite{Fann09}
\begin{align}
\tilde{H} C &= \tilde{S} C E , \: \: \: \tilde{H}_{i,j} = \int \psi_i^n (\vec{r})^\star \hat{H} \psi_j^n (\vec{r}) d \vec{r}, \\
 \tilde{S}_{i,j} &= \int \psi_i^n (\vec{r})^\star \psi_j^n (\vec{r}) d \vec{r} ,
\end{align} 
for $C$ and $E$. The new states $\phi_i^n$ with energies $E_i$ are obtained via
\begin{align}
\phi_i^n (\vec{r}) =  \sum\nolimits_{j} \psi_j^n (\vec{r}) C_{ij} .
\end{align}
We then apply the potential operator $U_\mathrm{HF,q}$ and perform the convolution with $G_\mathrm{BSH,i}$. This gives us a new set of states at iteration $n+1$:
\begin{align}
\phi_{i}^{n+1} &= -\alpha \: G_{BSH,i}  \star ( U_\mathrm{HF,q} \: \phi_i^{n} ) \\
                         &= - \alpha \left( - \Delta -  \alpha E_i  \right)^{-1} \: U_\mathrm{HF,q} \: \phi_{i}^{n} ,\nonumber\\
     U_\mathrm{HF,q}  &= U_q  - \nabla \cdot \left( b_1 \rho - b^{\prime}_1 \rho_q \right) \nabla + \mathrm{i} \vec{W}_q \cdot (\vec{\sigma} \times \nabla ). 
\end{align} 
To determine the convergence of the states we calculate the maximal L2-norm of the wave-function difference between two iterations: 
\begin{align}
\delta \psi &= (\delta \psi_i)_\mathrm{max} = \mathrm{max} \{ \delta \psi_0, ..., \delta \psi_{N_q} \} ,  \\
\delta \psi_i &=  \left( \int \left| \phi_i^{n+1} (\vec{r}) - \phi_i^{n} (\vec{r}) \right|^2 d \vec{r} \right)^{1/2} .
\label{convergence}
\end{align}
If $\delta \psi$ is smaller or equal to a given desired precision $\epsilon$, the iterations are considered as converged. Otherwise, the new single-particle states are calculated as averages of the old and new wavefunctions: 
\begin{align}
\psi_i^{n+1} = \chi \phi_i^{n+1} + (1 - \chi) \phi_i^{n}
\end{align} 
and the next iteration is performed. The averaging is a usual technique in HF calculations to stabilize the iterations. In our simulations we typically use $\chi = 0.4$. A larger value of $\chi$ leads to a faster convergence but might also allow the development of instabilities. Note that there are different iteration routines as well as convergence criteria \cite{Davies80, Reinhard82, Cusson85, Newton09} which might be more suitable for SHF calculations and will be tested in the future. For the present study we apply the same iteration steps and check for convergence as has been done in previous MADNESS studies \cite{Fann09}.
\section{Nuclear ground states}
\label{ground}
Before we apply our code to study nuclear pasta phases, we want to test its performance and accuracy by simulating the known ground states of several nuclei -  $^{16} \mathrm{O}$, $^{208} \mathrm{Pb}$, and $^{238} \mathrm{U}$. The first, $^{16} \mathrm{O}$, is a doubly magic nucleus with a well-known binding energy and is therefore a good first benchmark test of our code. Similarly, $^{208} \mathrm{Pb}$ is also doubly magic but with 13 times more nucleons than $^{16} \mathrm{O}$. Finally, $^{238} \mathrm{U}$ is a deformed nucleus and thereby a good test case for our code to find the nuclear ground state throughout several shape changes.\\
We calculate nuclei in (I) a large simulation space with free boundary conditions (bc) and no Jellium, and (II) a small box with periodic bc including the Jellium approximation. While the first is suitable for comparisons with experimental ground states, the second case has similar conditions as our nuclear pasta simulations. Furthermore, in addition to experimental binding energies \cite{Audi03}, we compare our results with the Skyrme HF code Sky3D \cite{Schuetrumpf13, Schuetrumpf14, Maruhn14, Schuetrumpf15}. For the setup of simulation (I) in M-SHF, we choose a box width of $L = 200 \: \mathrm{fm}$, while for (II) we apply $L = 24\: \mathrm{fm}$. The latter is used in Sky3D for simulations (I) and (II). Although $L = 24\: \mathrm{fm}$ is a relatively low value, the nuclear densities are $\lesssim 10^{-3}\: \mathrm{fm}^{-3}$ at the edges of the simulation space - even for large nuclei. However, to cross-check that the small simulation volume does not alter the nuclear ground states for simulation (I), we perform calculations with $L=48\:$fm and compare the total energies to $L=24\:$fm. Finally, to test the dependence on resolution in Sky3D, we vary the cell size of the computational grid. For MADNESS, we decrease the final truncation threshold $\epsilon$ and check the impact of Gaussian smoothing.\\
\newline
The single-particle states with spin-up and spin-down are initialized as harmonic oscillator states:
\begin{align}
\psi_{i,s} (\vec{r}) = x^j \: y^k \: z^l \: e^{\left( - r^2/ (2d)  \right) }, \:\:i = 0, ..., N_s - 1
\label{ho}
\end{align} 
where $j, \: k$ and $l$ are integers starting at 0, $N_s$ is the number of states with spin-up or spin-down, and $d = 0.625 \: A^{1/3}$. We use the Skyrme force SV-bas \cite{Kluepfel09} and calculate the binding energy from the energy components of the Skyrme density functional: 
\begin{align}
E_0 &= \frac{1}{2} \int  b_0 \: \rho^2 (\vec{r}) - b^{\prime}_0 \: \sum_q  \rho_q^2 (\vec{r}) \: d \vec{r} ,  
\label{energy0}\\
E_1 &=  \int  b_1 \rho (\vec{r}) \tau (\vec{r}) - b^{\prime}_1 \sum_q  \rho_q (\vec{r}) \tau_q (\vec{r})  \: d \vec{r} , 
\label{energy1}
\end{align}
\begin{align}
E_2 &=  \frac{1}{2} \int  b^{\prime}_2 \sum_q \rho_q (\vec{r}) \Delta \rho_q (\vec{r}) - b_2  \: \rho (\vec{r}) \Delta \rho (\vec{r}) \: d \vec{r} , 
\label{energy2}\\
E_3 &=  \frac{1}{3} \int b_3 \: \rho^{\alpha + 2} (\vec{r}) - b^{\prime}_3 \: \rho^{\alpha} (\vec{r}) \sum_q \rho_q^2 (\vec{r})  \: d \vec{r} , 
\label{energy3}
\end{align}
\begin{align}
E_4 = - \int  b_4 \: \rho (\vec{r}) \: \nabla \vec{J} (\vec{r}) + b^{\prime}_4 \sum_q \rho_q (\vec{r})  \: \nabla \vec{J} (\vec{r}) \:  d \vec{r} , 
\label{energy4}
\end{align}
the kinetic and Coulomb energies:
\begin{align}
E_\mathrm{kin} &= \sum_q \frac{\hbar^2}{2m} \int  \tau_q(\vec{r}) \: d \vec{r}, 
\label{energy_kin}\\
E_{C} &= \frac{1}{2} \int U_C  (\vec{r}) \rho_p (\vec{r}) \: d \vec{r}    .
\label{energy_coulomb}
\end{align}
For the latter, the total charge potential $U_C$ and $\rho_C$ are given by eq.(\ref{VC_rhoC}) for simulation (II) and $U_C = U_{pp}$ for (I). The total binding energy and binding energy per nucleon are then given by:
\begin{align}
E_\mathrm{total} = E_\mathrm{kin} + E_C + E_0 + E_1 + E_2 + E_3 + E_4
\end{align}
and $E_\mathrm{bind} = E_\mathrm{total} / \mathrm{A}$, respectively. As in Sky3D, we do not consider energy contributions beyond the mean-field approximation, for example a center-of-mass correction \cite{Maruhn14}. The latter decreases with higher mass number as $\sim 1/A$ and should therefore have only a small contributions in simulations of heavy nuclei and nuclear pasta. However, for light nuclei, the lack of a center-of-mass correction might lead to noticeable deviations from experimentally obtained binding energies.
Results of the simulations are given in tables \ref{table_O16} - \ref{table_U238}. Here, for Sky3D and the MADNESS SHF code (abbreviated as M-SHF), we give the box length $L$ together with the simulation setup (I) or (II). The resolution of the simulation is given by the grid cell size $\Delta x$ for Sky3D and truncation threshold $\epsilon$ for MADNESS. The binding energy and total binding energy, $E_\mathrm{bind}$ and $E_\mathrm{total}$, respectively, are followed by the different energy components as described in eq.(\ref{energy0}) - eq.(\ref{energy_coulomb}). All energies are given in units of MeV. For the nuclear simulations with M-SHF, we typically start out with $\epsilon = 10^{-4}$ and Gaussian smoothing using $\sigma=0.25$. When
\begin{equation}
\delta \psi \sim A \times \epsilon
\label{trunc_crit}
\end{equation}
we decrease the truncation threshold to $\epsilon = 10^{-5}$. At this point, we continue with three versions of the simulation. The first version still contains the Gaussian smoothing while we remove it in the second one. Both simulations are evolved until eq.(\ref{trunc_crit}) is fulfilled for the new $\epsilon$. At this point, the calculations are stopped. We can then compare the impact of the smoothing on the energies and the nuclear configuration. In a third simulation, we continue the simulation without smoothing and decrease $\epsilon$ by factors of 10 according to eq.(\ref{trunc_crit}) while $\epsilon \geq 10^{-7}$. The simulations are run on the high performance computer center BigRed\:II at Indiana University and the EOS cluster at the Oak Ridge Leadership Computing Facility where we use nodes with 16 cores. As our code is not yet optimized for speed we do not give specific runtime numbers at this point but provide typical order-of-magnitude iteration counts and computational times for each nucleus and pasta simulations. Note that the time-independent calculations with Sky3D are not MPI parallelized and run at BigRed\:II on one node with 32 cores. As Sky3D represents wavefunctions and performs calculations on a fixed grid, simulations times for a given number states scale roughly by a factor of eight when $L$ is increased by a factor of two or $\Delta x$ decreased to half its size. Due to the adaptive refinement, MADNESS simulation times depend mostly on the truncation threshold $\epsilon$ (again assuming a fixed number of wavefunctions).  
\subsection{$^{16} \mathrm{O}$ nucleus}
\begin{figure}
\includegraphics[width = 0.45\textwidth]{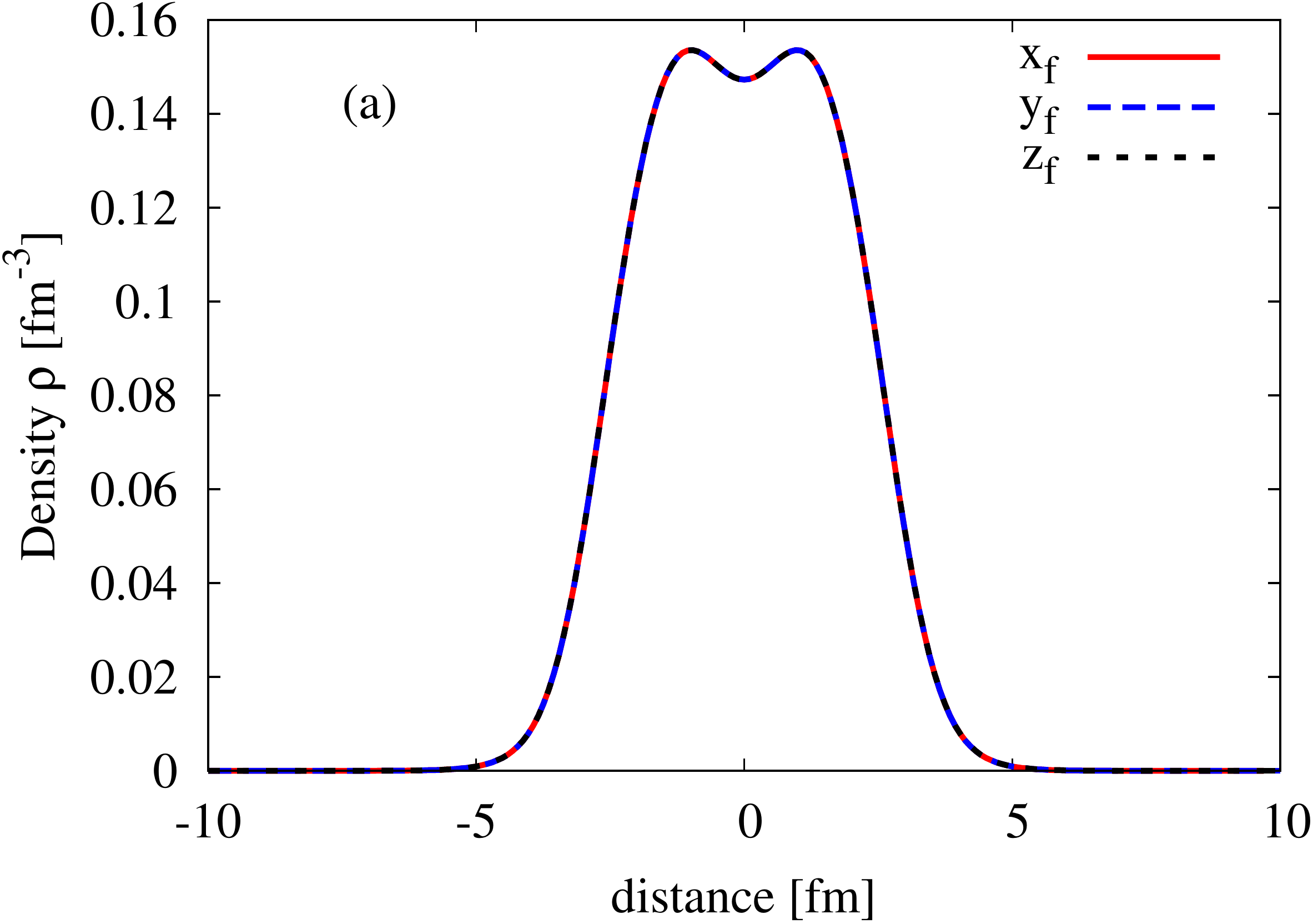}
\includegraphics[width = 0.45\textwidth]{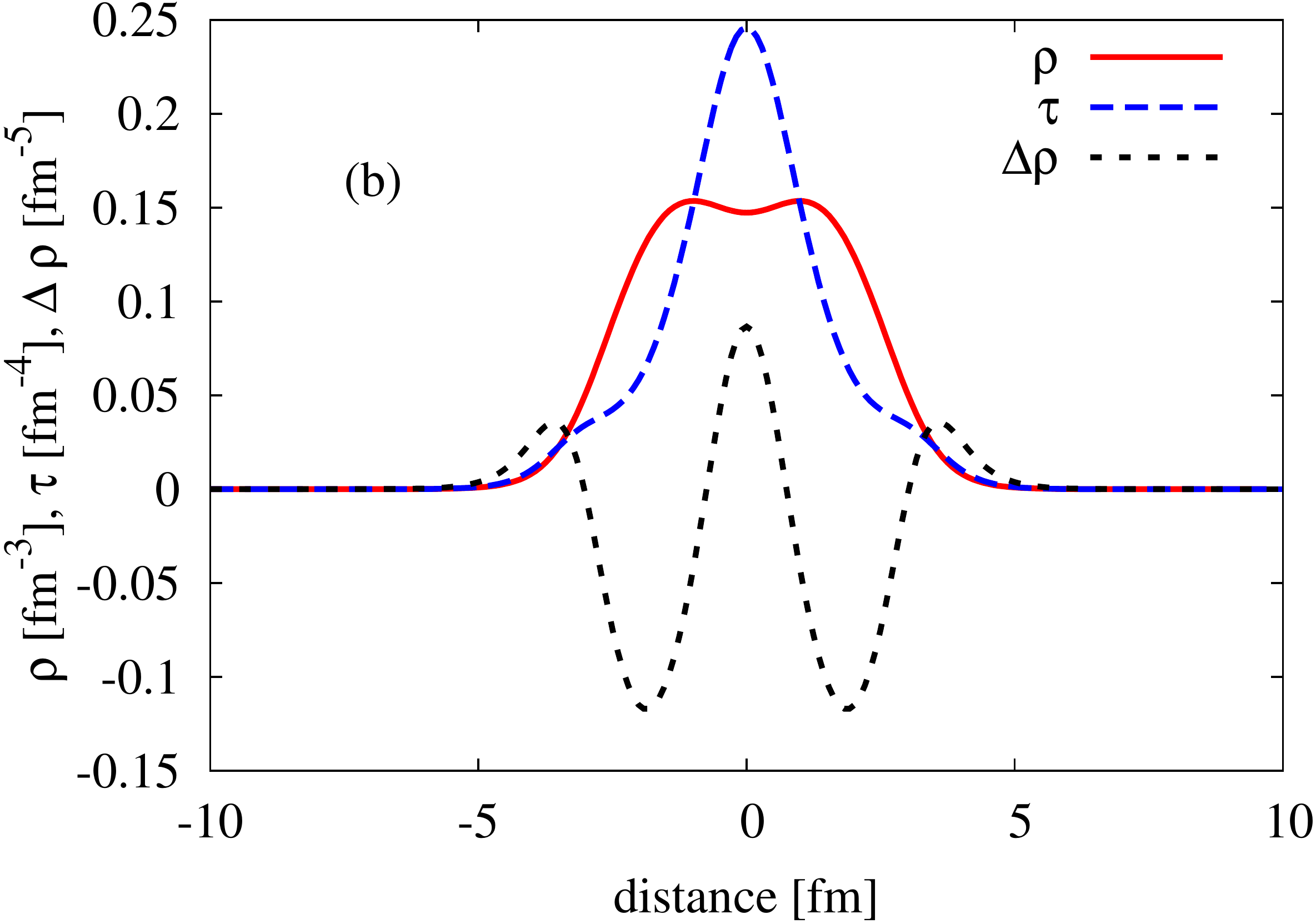}
\caption{(a) Number density profiles of $^{16}\mathrm{O}$ nucleus ground state, as obtained with our MADNESS code, taken along the x, y, and z-axis. The profiles show that the resulting nucleus is spherically symmetric. (b) Number density, kinetic density $\tau$ and laplacian of the number density $\Delta \rho$ along the x-axis.}
\label{pic_O16}
\end{figure}

Our results for the $^{16} \mathrm{O}$ nucleus are given in table \ref{table_O16} for the Sky3D and M-SHF simulations. We first discuss the results for setup (I). Figure \ref{pic_O16}\:(a) shows the number density profiles of the converged $^{16}\mathrm{O}$ nucleus for M-SHF with $\epsilon = 10^{-7}$ along the three axes and $-10\:$fm $\leq x,y,z \leq 10\:$fm. Due to the spherical shape of the nucleus, the profiles overlap exactly. The kinetic density and laplacian of the total density $\rho$ are shown in Fig.\:\ref{pic_O16}\:(b) along the x-axis, together with the corresponding density profile. Despite the absence of Gaussian smoothing, the profiles show no discontinuities or irregularities and, due to the spherical symmetry of the nucleus, are identical along all axes. In table \ref{table_O16}, we find a clear difference between energies for simulations with Gaussian smoothing (marked with a $\star$ in the resolution column) and without, whereas all components are affected. The difference in $E_\mathrm{total}$ between a simulation using smoothing and without is $| \Delta E_\mathrm{total}|  \sim 1.458 \times 10^{-2} \: | E_\mathrm{total}| $ for $\epsilon = 10^{-5}$, whereas this value will depend on the size of $\sigma$. On the other hand, when comparing simulations without smoothing but with different truncation thresholds, $\epsilon = 10^{-5}$ and $\epsilon = 10^{-7}$, we do not find any noticeable differences for up to 7 decimals in $E_\mathrm{total}$ (6 decimals in the table). \\
For Sky3D, changing the box size from $L=48\:$fm to $L=24\:$fm leads to a difference in total energy of only $|\Delta E_\mathrm{total}| \sim 1.20 \times 10^{-4} \: | E_\mathrm{total}| $. This is smaller than the change due to a decrease in cell size from $\Delta x = 1\:$fm to $0.5\:$fm which results in $|\Delta E_\mathrm{total}| \sim 3.86 \times 10^{-4} \: |E_\mathrm{total}|$. A further reduction to $\Delta x = 0.25\:$fm has a negligible effect which implies that, at least for $^{16} \mathrm{O}$, $\Delta x = 0.5\:$fm is sufficient to capture the correct energy values.\\ 
\begin{table*}
\centering
\begin{tabularx}{\textwidth}{@{}  l c c c Y Y Y Y Y Y Y Y Y@{}  }                    
\hline
   & L [fm] &  sim. & resol. & E$_\mathrm{bind}$  &  E$_\mathrm{total}$  & E$_\mathrm{kin}$  & E$_0$  & E$_1$  & E$_2$  & E$_3$  & E$_4$  & E$_\mathrm{C}$  \\
\hline 
\hline 
Sky3D   & 48   &(I)      & 1.0 fm           & -7.290               & -116.643 & 234.538 & -976.171  & 12.689 & 43.471  & 556.049 & -0.747 & 13.542  \\
Sky3D   & 24   &(I)      & 1.0 fm           & -7.291               & -116.657 & 234.537 & -976.163  & 12.689 & 43.470  & 556.044 & -0.747 & 13.542  \\
Sky3D   & 24   &(I)      & 0.5 fm           & -7.288               & -116.612 & 234.443 & -976.109  & 12.687 & 43.499  & 556.086 & -0.752 & 13.535  \\
Sky3D   & 24   &(I)      & 0.25 fm         & \textbf{-7.288}   & -116.613 & 234.443 & -976.113  & 12.687 & 43.499  & 556.088 & -0.752 & 13.534  \\
\hline  
M-SHF  & 200 &(I)$\star$      & $10^{-5}$  & -7.396               & -118.336 & 237.165 & -992.391   & 12.801 & 43.528 & 567.719 & -0.765 & 13.606 \\ 
M-SHF  & 200 &(I)                 & $10^{-5}$  & -7.288               & -116.611 & 234.444 & -976.114   & 12.688 & 43.499 & 558.088 & -0.752 & 13.535 \\ 
M-SHF  & 200 &(I)                 & $10^{-7}$  & \textbf{-7.288}   & -116.611 & 234.444 & -976.114   & 12.688 & 43.499 & 556.088 & -0.752 & 13.535 \\
\hline
\hline
Sky3D   & 24   &(II)     & 1.0 fm          & -7.626               & -122.010 & 234.606 & -976.537   & 12.697 & 43.497 & 556.312 & -0.747 &  8.192  \\ 
Sky3D   & 24   &(II)     & 0.5 fm          & -7.622               & -121.958 & 234.504 & -976.439   & 12.694 & 43.522 & 556.322 & -0.752 &  8.191  \\
Sky3D   & 24   &(II)     & 0.25 fm        & \textbf{-7.622}   & -121.958 & 234.504 & -976.438   & 12.694 & 43.522 & 556.321 & -0.752 &  8.191  \\
\hline          
M-SHF  & 24   &(II)$\star$     & $10^{-5}$    & -7.730               & -123.683 & 237.222 & -992.699   & 12.807 & 43.549 & 567.940 & -0.764 & 8.262    \\  
M-SHF  & 24   &(II)                & $10^{-5}$    & -7.622               & -121.957 & 234.505 & -976.440   & 12.694 & 43.522 & 556.321 & -0.752 & 8.192     \\ 
M-SHF  & 24   &(II)                & $10^{-7}$    & \textbf{-7.622}   & -121.957 & 234.505 & -976.440   & 12.694 & 43.522 & 558.321 & -0.752 & 8.192    \\  
\hline
\end{tabularx}
\caption{Parameters and energies for Sky3D and M-SHF simulations of the $^{16} \mathrm{O}$ nucleus. Simulations with free boundary conditions are marked by (I) while periodic boundary conditions with the jellium approximation are given by (II). The simulation box size is given by its length $L$. The resolution is defined as the grid cell size for Sky3D and truncation threshold for M-SHF. Simulations that apply Gaussian smoothing are marked by a $\star$. The binding energy per baryon $E_\mathrm{bind}$, total energy $E_\mathrm{total}$ and different energy components: $E_\mathrm{kin}$, $E_0$ - $E_4$, and $E_C$ (see eq.(\ref{energy0}) - eq.(\ref{energy_coulomb})) are given in MeV. Binding energies for simulations that were performed with the highest precisions are marked by bold font.}
\label{table_O16}
\end{table*}
Regarding their sensitivity to resolution, results for simulations (II) are very similar to the ones of setup (I). From table \ref{table_O16}, we see again a difference in the total energy for Sky3D simulations when changing the cell size from $\Delta x = 1\:$fm to $0.5\:$fm, and no effects for a further decrease to $\Delta x = 0.25\:$fm. Similarly, for M-SHF, Gaussian smoothing leads to $| \Delta E_\mathrm{total}|  \sim 3.51 \times 10^{-3} \: |E_\mathrm{total}|$, while a decrease in truncation threshold does not have any visible effects. However, for all simulations, the absolute values of the total energy are higher than in setups with free bc. This is due to the smaller Coulomb energy as a consequence of the Jellium. 
\begin{figure}
\begin{center}
\includegraphics[width = 0.45\textwidth]{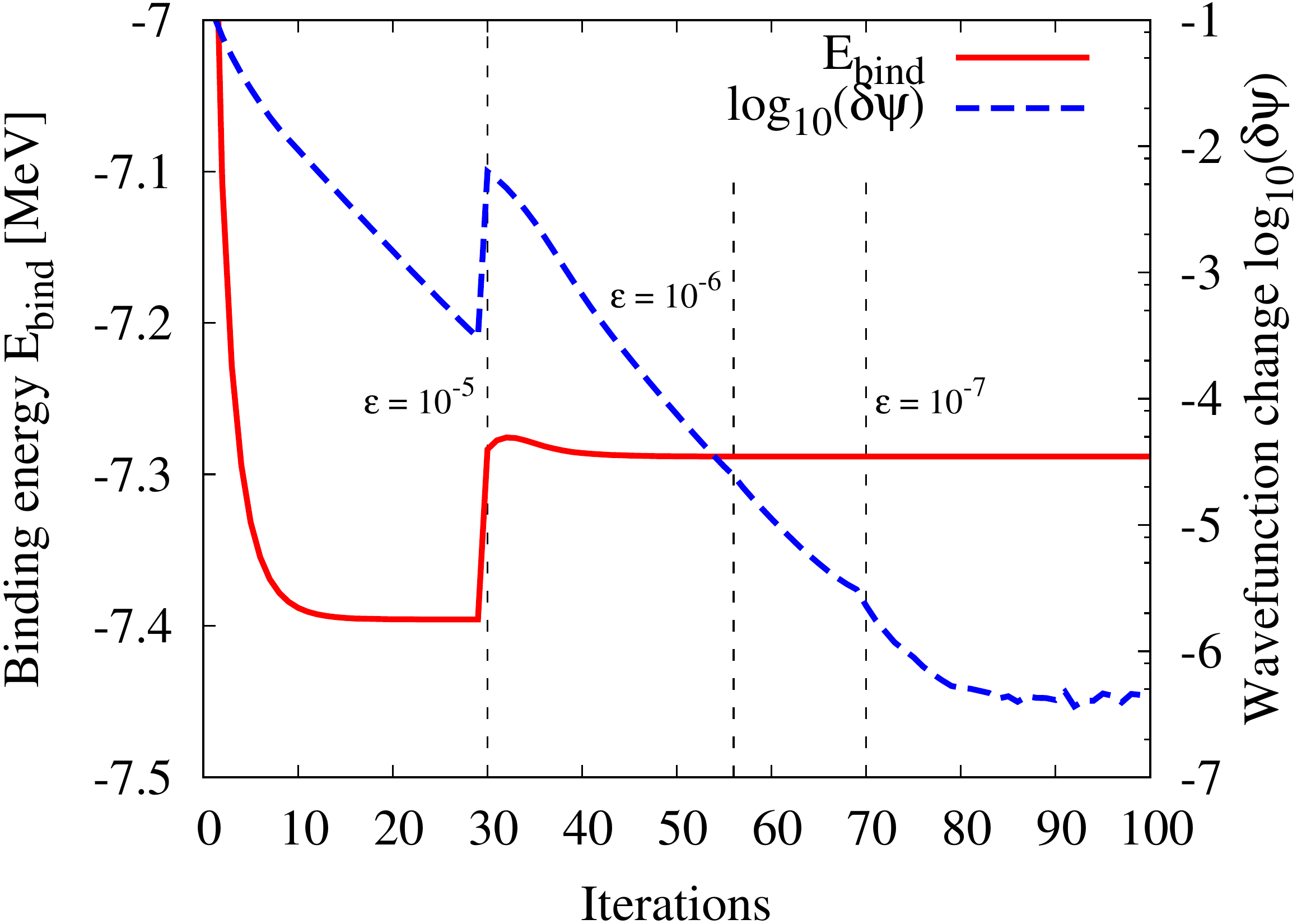}
\caption{Evolution of the maximum error and binding energy with iterations of the $^{16}\mathrm{O}$ calculation.}
\label{O_log}
\end{center}
\end{figure}
The latter reduces the total electric charge density and thereby the Coulomb potential. Although M-SHF requires less than hundred iterations until convergence while Sky3D uses several hundred steps, both codes are very fast and require less than one hour on one node. Figure \ref{O_log} shows the maximum error $\delta \psi$ and the binding energy per particle for M-SHF as they evolve with iterations for setup (I) and a final truncation threshold of $\epsilon = 10^{-7}$. The vertical dashed lines mark the reduction of the truncation threshold to the new values as given in the figure. The large jump in $\delta \psi$ and $E_\mathrm{bind}$ at iteration $\sim 30$ is due to the removal of Gaussian smoothing. After that, we see that the binding energy does not change much until the calculation is converged and $\delta \psi$ becomes constant. The MADNESS calculations of $^{16} \mathrm{O}$ and the Sky3D results are in good agreement with each other for simulations (I) and (II) and differ by only $| \Delta E_\mathrm{total} |  \sim 0.002\:$MeV when using the highest discussed resolutions. The large deviation from the experimental binding energy of $E_\mathrm{exp} \sim -7.976\:$MeV \cite{Audi03} originates in the applied Skyrme force Sv-bas and, as previously mentioned, could be partially attributed to the absence of the center-of-mass correction.
\subsection{$^{208} \mathrm{Pb}$ nucleus}
\begin{figure*}
\begin{center}
\includegraphics[width = 0.45\textwidth]{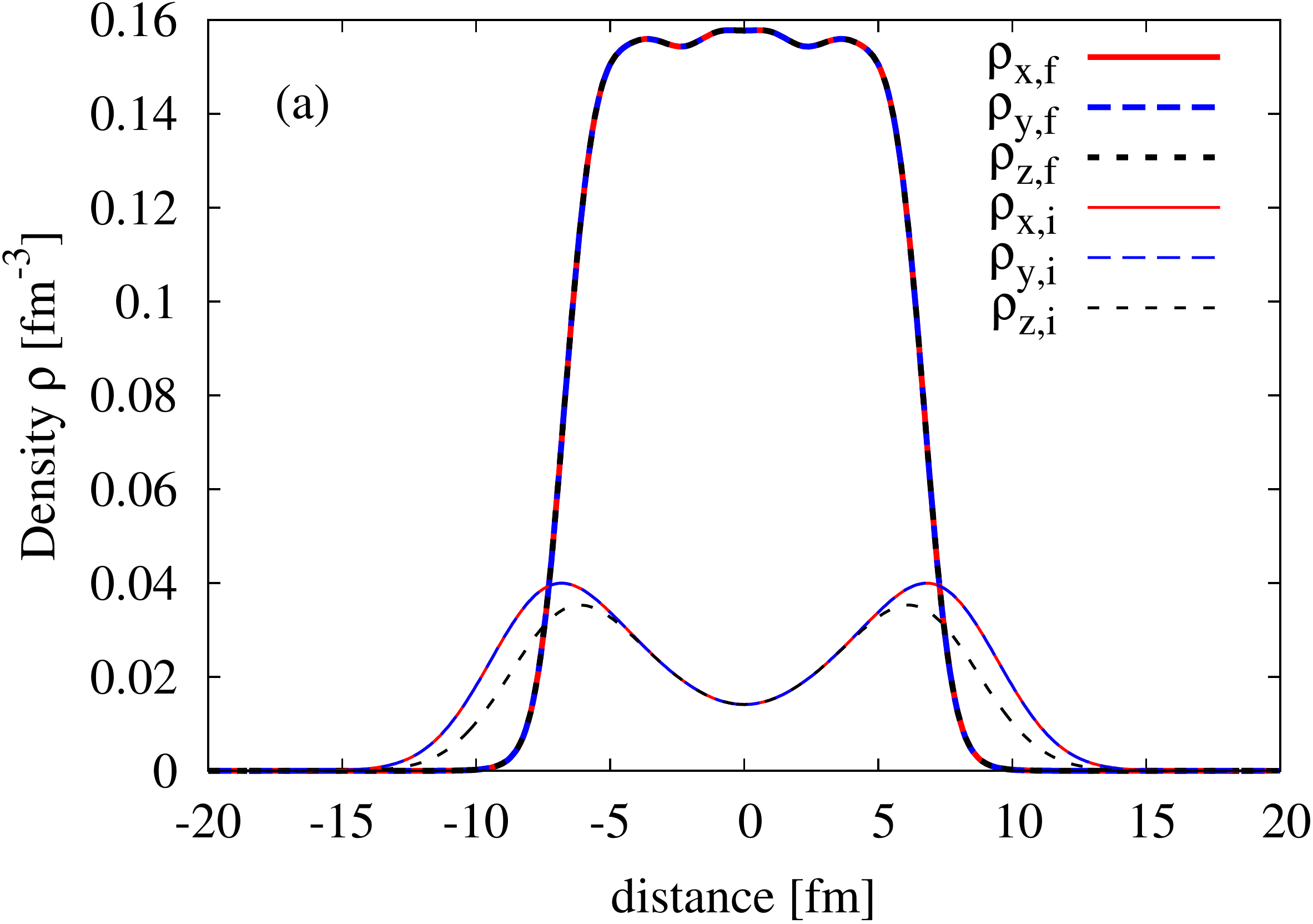}
\includegraphics[width = 0.45\textwidth]{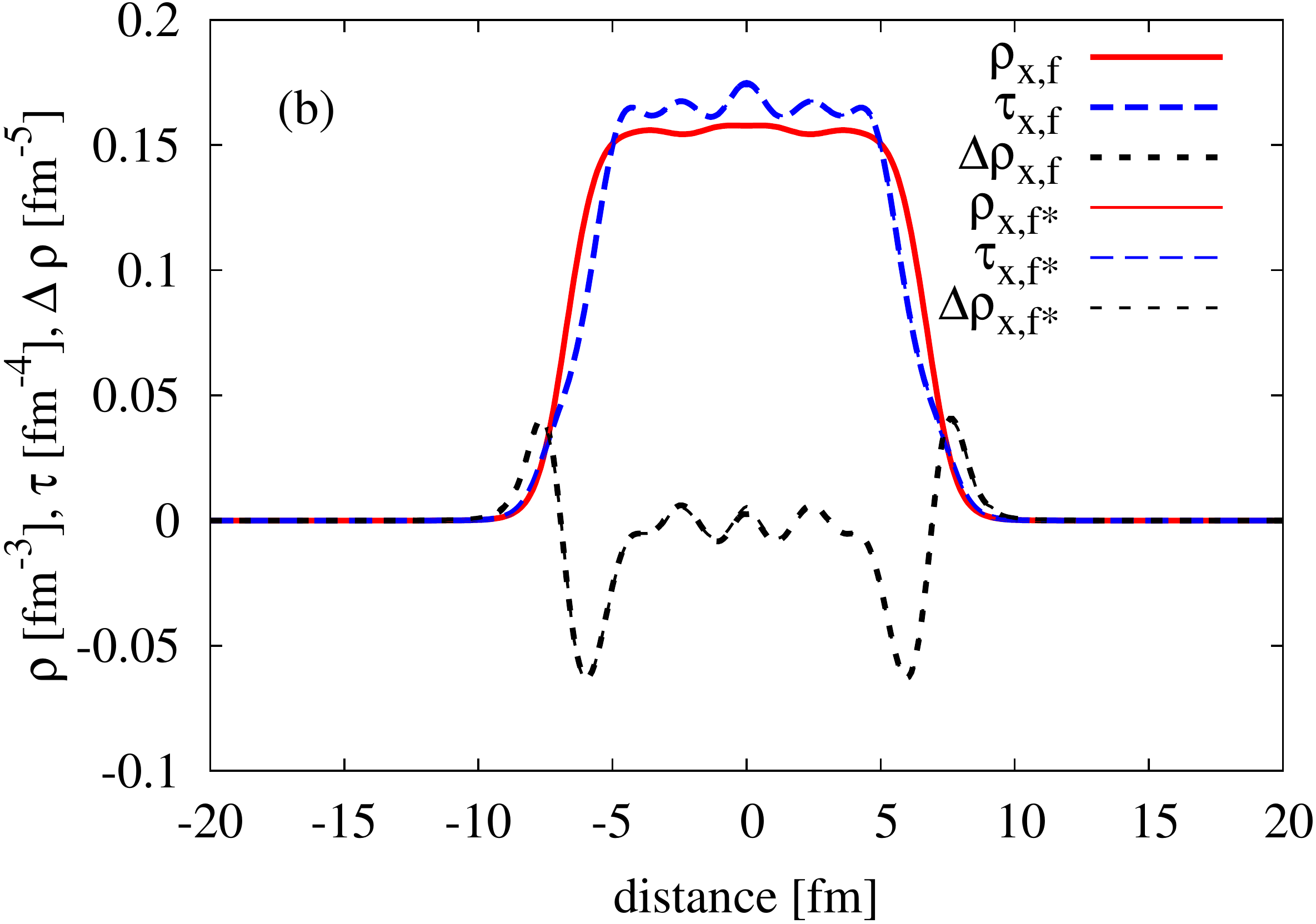}
\caption{(a) Number density profiles of the $^{208}\mathrm{Pb}$ along the x, y, and z-axis in the initial state (subscript $i$) and at convergence (subscript $f$). Subfigure (b) shows the number density, kinetic density $\tau$ and laplacian of the number density $\Delta \rho$ at iteration 280.}
\label{Pb208}
\end{center}
\end{figure*}

Next, we discuss simulations of the $^{208}\mathrm{Pb}$ nucleus with Sky3D and M-SHF. The resulting energies are summarized in table \ref{table_Pb208} where the structure of the table is as for $^{16}\mathrm{O}$. Interestingly, for Sky3D, the difference in the total energies of $^{208}\mathrm{Pb}$ between $L=24\:$fm and $L=48\:$fm in simulation (I) is the same as for $^{16}\mathrm{O}$, namely $| \Delta E_\mathrm{total} |  \sim 0.014\:$MeV. In comparison to the total energy of $^{208}\mathrm{Pb}$ it is of course only $\sim 8.53 \times 10^{-6} | E_\mathrm{total} |$. When increasing the resolution by setting $\Delta x = 0.5\:$fm, the total energy changes by about $\sim 0.058\:$MeV or $\sim 3.56 \times 10^{-5} E_\mathrm{total}$. As before, it seems that the change in energy due to the simulation volume is smaller than the one caused by an increase in resolution.  Setting $\Delta x = 0.25\:$fm results in $| \Delta E_\mathrm{total}|  \sim 0.003\:$MeV which is negligible in comparison to the total energy. As for $^{16}\mathrm{O}$, we can conclude that a resolution of $\Delta x = 0.5\:$fm is sufficient to reproduce the ground state of $^{208}\mathrm{Pb}$ and a cell size of $L=24\:$fm does not lead to large finite-size effects. For $^{208}\mathrm{Pb}$ in simulation (II) and the $^{238}\mathrm{U}$ calculations we will therefore only test $L=24\:$fm and $\Delta x \geq 0.5\:$fm. As we will see, the energetic differences between simulations with $\Delta x = 0.5\:$fm and $\Delta x = 1.0\:$fm are very small.\\ 
\begin{table*}
\centering
\begin{tabularx}{\textwidth}{@{}  l c c c Y Y Y Y Y Y Y c c @{}  }                    
\hline
     & L [fm]&  sim. & resol. & E$_\mathrm{bind}$ &  E$_\mathrm{total}$ & E$_\mathrm{kin}$ & E$_0$ & E$_1$ & E$_2$ & E$_3$ & E$_4$ & E$_\mathrm{C}$ \\
\hline 
\hline 
Sky3D   & 48  &(I)   & 1.0 fm            &-7.842         & -1631.027 & 3920.428 & -17379.720  & 285.346 & 240.119 & 10591.390 & -86.064 & 798.492  \\
Sky3D   & 24   &(I)   & 1.0 fm           &-7.841         & -1631.013 & 3920.767 & -17380.820  & 285.356 & 240.123 & 10591.170 & -86.059 & 798.489  \\ 
Sky3D   & 24   &(I)   & 0.5 fm           &-7.841         & -1630.955 & 3920.585 & -17380.620  & 285.356 & 240.101 & 10591.220 & -86.067 & 798.465  \\
Sky3D   & 24   &(I)   & 0.25 fm         &\textbf{-7.841}         & -1630.958 & 3920.587 & -17380.640  & 285.357 & 240.102 & 10591.240 & -86.067 & 798.462  \\
\hline
M-SHF  & 200 &(I)$\star$   & $10^{-5} $ &-7.845          &  -1631.834 & 3923.895 & -17397.536  & 285.436 & 239.842 & 10603.944 & -86.106 & 798.692  \\
M-SHF  & 200 &(I)   & $10^{-5} $  &-7.837         & -1630.160 & 3920.557 & -17376.751  & 285.309 & 240.030 & 10588.270 & -86.045 & 798.469  \\
M-SHF  & 200 &(I)   & $10^{-7} $  &\textbf{-7.837}         & -1630.162 & 3920.541 & -17376.675  & 285.309 & 240.021 & 10588.219 & -86.047 & 798.469   \\
\hline
\hline
Sky3D   & 24  &(II)   & 1.0 fm           &-10.385       & -2160.078 & 3940.437 & -17519.980 & 289.937 & 242.621 & 10701.650 & -87.118 & 272.412   \\ 
Sky3D   & 24   &(II)   & 0.5 fm           &\textbf{-10.385}       & -2159.991 & 3940.278 & -17520.010 & 289.943 & 242.615 & 10701.890 & -87.127 & 272.419  \\
\hline
M-SHF  & 24   &(II)$\star$    & $10^{-5}$  &-10.389          &  -2160.857 & 3943.894 & -17537.870  & 290.025 & 242.342 & 10715.300 & -87.162 & 272.616  \\
M-SHF  & 24   &(II)    & $10^{-5} $  &-10.381          &  -2159.163 & 3940.599 & -17517.276  & 289.908 & 242.549 & 10699.733 & -87.100 & 272.424  \\
M-SHF  & 24   &(II)    & $10^{-7} $  &\textbf{-10.381}          &  -2159.161 & 3940.577 & -17517.170  & 289.907 & 242.540 & 10699.663 & -87.101 & 272.424  \\
\hline  
\end{tabularx}
\caption{Parameters and energies for Sky3D and M-SHF simulations of the $^{208}$Pb nucleus. The table setup is the same as in table \ref{table_O16}. The experimental binding energy for $^{208}$Pb nucleus is E$_\mathrm{bind} \sim -7.867\:$MeV \cite{Audi03}.}
\label{table_Pb208}
\end{table*}
For M-SHF, we apply again Gaussian smoothing with $\sigma=0.25$ in the beginning of the simulation when we initialize the wavefunctions as harmonic oscillator states and truncate with $\epsilon = 10^{-4}$. Once eq.(\ref{trunc_crit}) is fulfilled, we continue with three simulations as described in the previous section. Table \ref{table_Pb208} shows that Gaussian smoothing affects again all energy terms, leading to a difference in $E_\mathrm{total}$ of $\sim 1.674\:$MeV or $\sim 1.02 \times 10^{-3} | E_\mathrm{total}| $. In contrast, reducing $\epsilon$ from $10^{-5}$ to $10^{-7}$ results in a small change of the total energy of only $\sim 0.002\:$MeV or $1.22 \times 10^{-6} | E_\mathrm{total}|$. This suggests that a final value of $\epsilon = 10^{-5}$ is sufficient to reproduce the nuclear ground state energies.\\
\newline
Figure \ref{Pb208}(a) shows the $x$, $y$ and $z$ profiles of the initial and final total density $\rho$ in simulation (I) for $\epsilon = 10^{-7}$ and $-20\:\mathrm{fm} \leq x,y,z \leq 20\:$fm. We can see that while the initial density distribution is slightly flatter in the $z$-direction, the final shape is spherically symmetric. The $x$-profile of the total density is again shown in Fig.\ref{Pb208}(b) together with the profiles for $\tau$ and $\Delta \rho$. 
\begin{figure}
\begin{center}
\includegraphics[width = 0.48\textwidth]{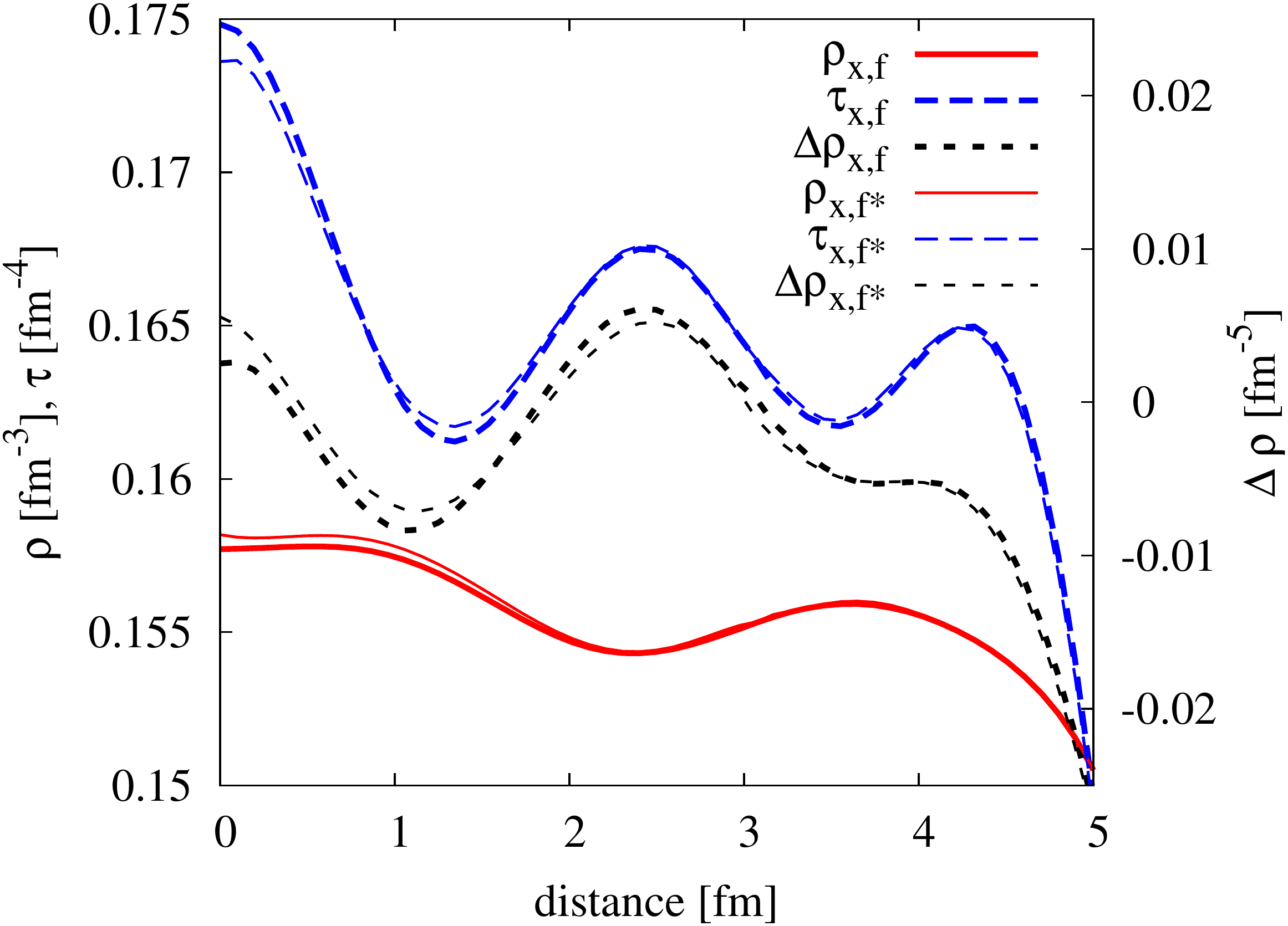}
\caption{Zoom of Fig.\ref{Pb208}(b) showing the $\rho$, $\tau$ and $\Delta \rho$ profiles along x for the $^{208}\mathrm{Pb}$ simulation (I) with smoothing and $\epsilon = 10^{-5}$ (marked by a $\star$) and $\epsilon = 10^{-7}$ without smoothing. While the oscillation patterns of the three quantities are similar with and without smoothing, small differences are visible in the amplitudes.}
\label{Pb208_sc_2}
\end{center}
\end{figure}
We also plot the corresponding final quantities for simulation (I) with $\epsilon = 10^{-5}$ and Gaussian smoothing. Differences in $\rho$ and $\tau$ are not visible while the $\Delta \rho$ profiles show small deviations around $x \sim 0\:$fm. For better comparison, we zoom into the nucleus and compare the profiles again in Fig.\ref{Pb208_sc_2}. Small deviations in the oscillation amplitudes of all three quantities can be seen. However, despite these differences, the oscillation pattern are very similar. As a consequence, while the application of Gaussian smoothing impacts the energies, the effects on the shape of a nuclear configuration might be small.\\     
As for $^{16}\mathrm{O}$, the sensitivity of the periodic simulations for $^{208}\mathrm{Pb}$ is similar to the free bc studies. For Sky3D, increasing the resolution by changing $\Delta x$ from $1.0\:$fm to $0.5\:$fm, results in $| \Delta E_\mathrm{total} | \sim 0.087\:\mathrm{MeV} \sim 4.03 \times 10^{-5} \: |E_\mathrm{total}|$. With $| \Delta E_\mathrm{total}| \sim 1.694\:\mathrm{MeV} \sim 7.84 \times 10^{-4} \: |E_\mathrm{total}|$, Gaussian smoothing in M-SHF has again a larger impact on the energies than reducing the truncation threshold which results in $|\Delta E_\mathrm{total}| \sim 0.002\: \mathrm{MeV}$. \\
In general, both codes agree well. The deviations between numerical results for $E_\mathrm{bind}$ in simulation (I) and the experimental value of $\sim -7.867\:$MeV are due to the applied Skyrme force SV-bas. The higher energies in simulations with periodic boundary conditions are again due to the inclusion of Jellium. Although, the difference in total energy between both codes results in only $\sim 0.8\:$MeV or $\sim 0.004\:$MeV per baryon, it is notable and should be understood. As can be seen from tables \ref{table_O16} and \ref{table_Pb208}, the difference $\Delta E_\mathrm{total}$ between Sky3D and M-SHF seems to scale with the number of states and is most pronounced in the $E_0$ and $E_3$ terms which are both functions of the baryon densities and have large absolute values. Our tests of $L$, $\Delta x$ and $\epsilon$ did not reveal any sensitivities of the results that would be large enough to account for the energy difference. Furthermore, since we see the same behavior for simulations (I) and (II), we can assume that it is not an effect of specific boundary conditions. With that, more cross-checks have to be performed between both codes in the future.\\
\newline
Due to the larger number of wavefunctions, calculations of $^{208}\mathrm{Pb}$ take longer than for $^{16}\mathrm{O}$. Simulations with Sky3D, required ca. 3.5 hours and about 1000 iterations on one node for $\Delta x = 1.0\:$fm. With M-SHF, simulations of $^{208}\mathrm{Pb}$ were performed on 8 CPU nodes. Gaussian smoothing has only a minor impact on the simulation time while setting the truncation threshold to lower values increases the latter. However, in all cases, calculations reach the ground state within several hours and $\leq 300$ iterations. The binding energy and $\log(\delta \psi)$ as 
\begin{figure}
\begin{center}
\includegraphics[width = 0.45\textwidth]{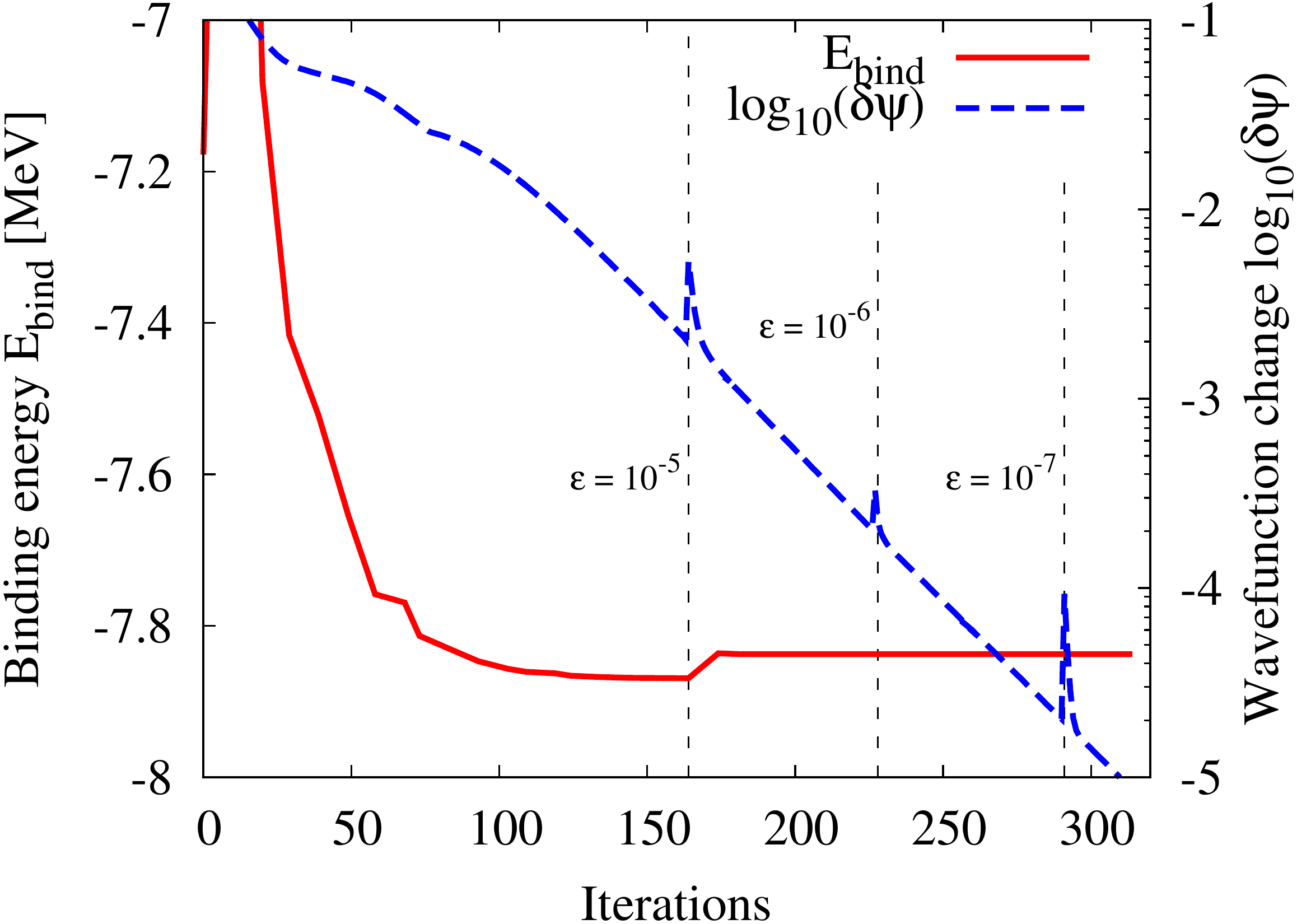}
\caption{Evolution of the maximum error and binding energy with iterations of the $^{208}\mathrm{Pb}$ calculation.}
\label{Pb208_log}
\end{center}
\end{figure}
functions of iteration number for simulation (I) and final truncation threshold $\epsilon = 10^{-7}$ are plotted in Fig.\:\ref{Pb208_log}. The binding energy is calculated every 10 iterations and the dashed vertical lines mark again the reduction of $\epsilon$. We can see three peaks in $\delta \psi$ which are caused by the decrease of the truncation threshold according to eq.(\ref{trunc_crit}). The first decrease is accompanied by the removal of Gaussian smoothing which also leads to a jump in the binding energy. However, except for the three discontinuities, the value of $\delta \psi$ gradually decreases, which indicates that the initialization via harmonic oscillator states is a good initial guess and does not require any major shape changes of the nucleus. 
\subsection{$^{238} \mathrm{U}$ nucleus}
\begin{figure*}
\begin{center}
\includegraphics[width = 0.45\textwidth]{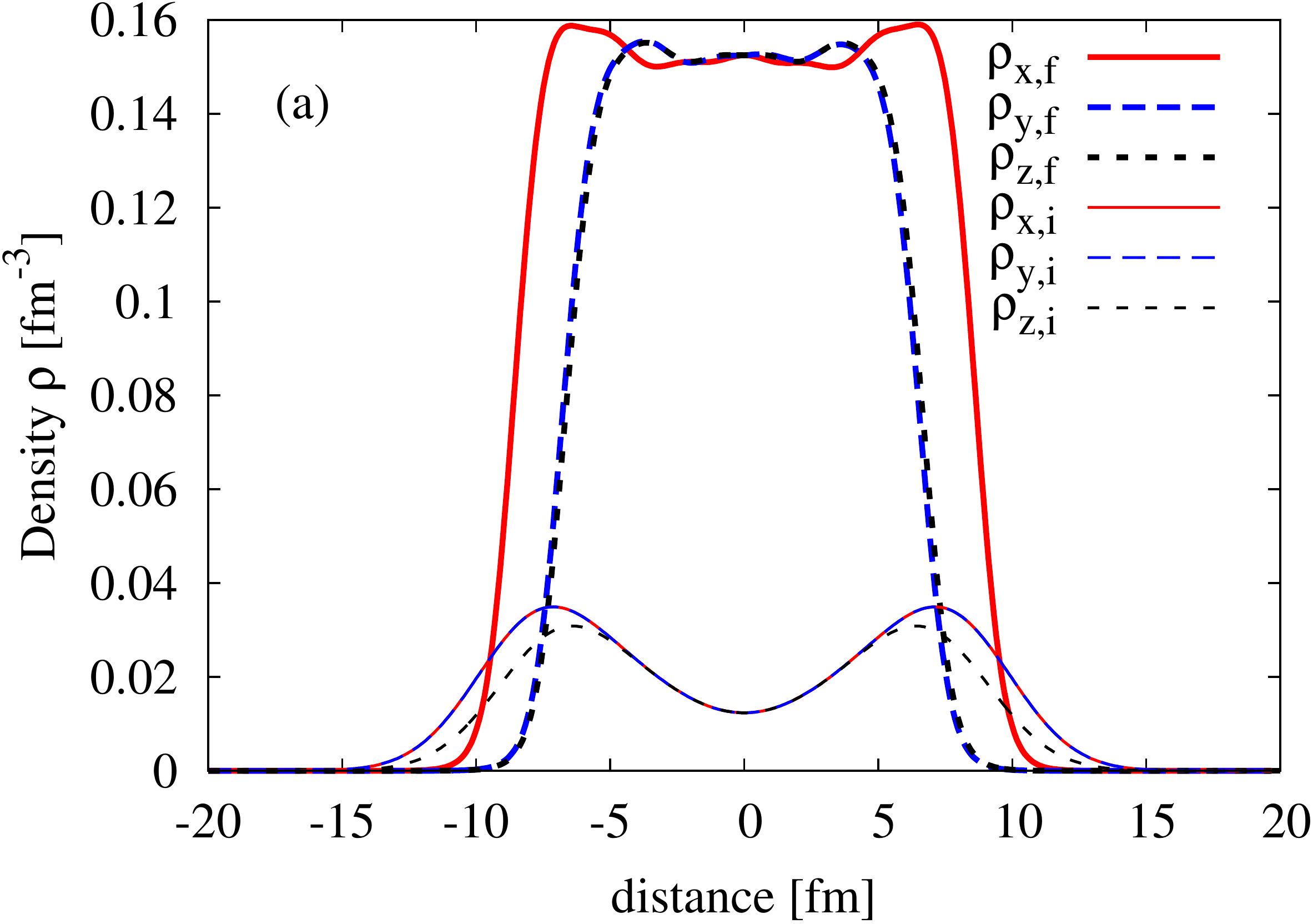}
\includegraphics[width = 0.45\textwidth]{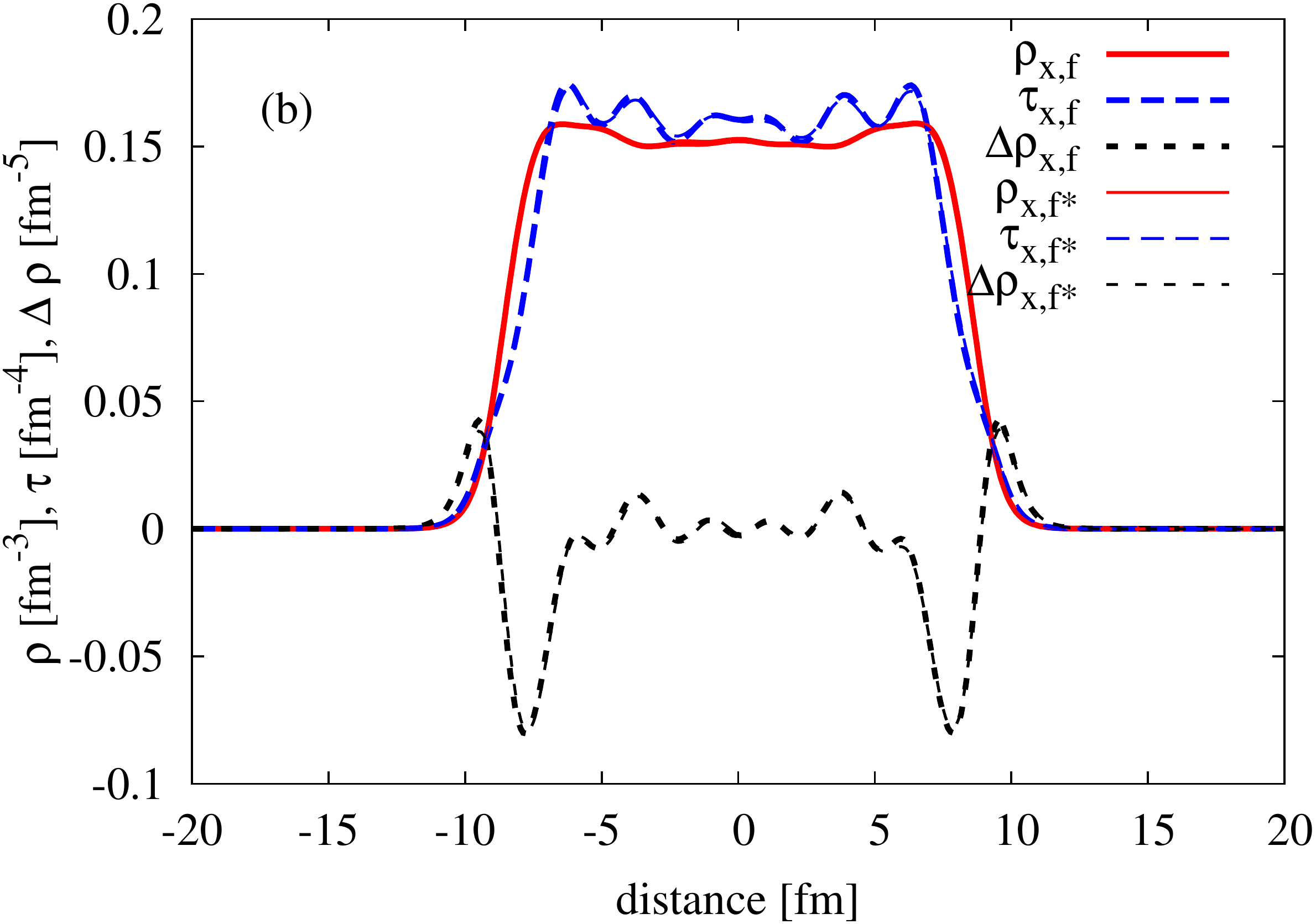}
\caption{(a) Number density profiles of the $^{238}\mathrm{U}$ along the x, y, and z-axis at convergence (subscript $f$) and the beginning of the iterations (subscript $i$). Subfigure (b) shows the number density, kinetic density $\tau$ and laplacian of the number density $\Delta \rho$ at iteration 2800.}
\label{U238}
\end{center}
\end{figure*}

\begin{figure}
\begin{center}
\includegraphics[width = 0.48\textwidth]{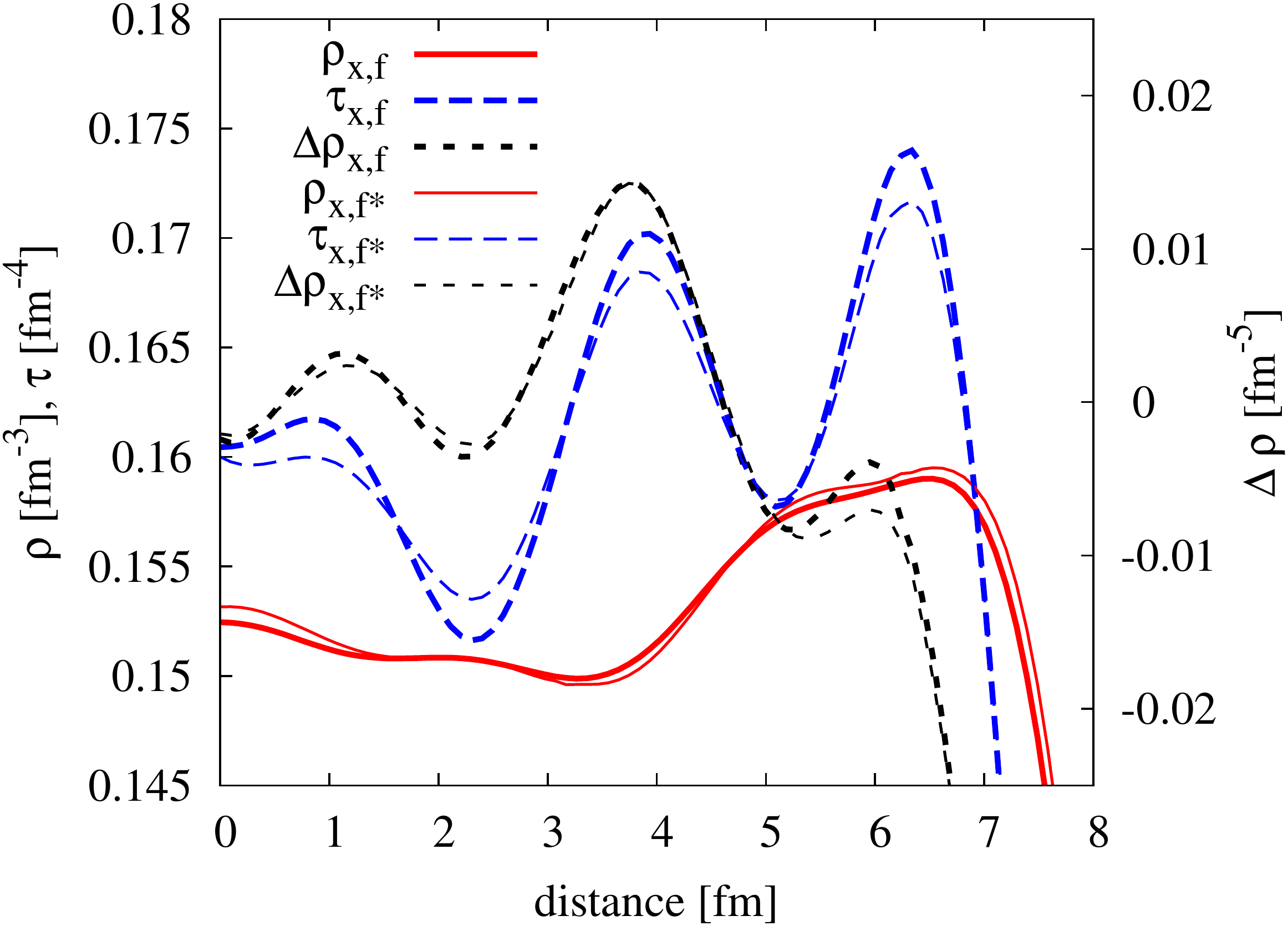}
\caption{Zoom of Fig.\ref{U238}(b) showing the $\rho$, $\tau$ and $\Delta \rho$ profiles along x for the $^{238}\mathrm{U}$ simulation (I) with smoothing and $\epsilon = 10^{-5}$ (marked by a $\star$) and $\epsilon = 10^{-7}$ without smoothing.}
\label{U238_sc_2}
\end{center}
\end{figure}

\begin{figure}
\begin{center}
\includegraphics[width = 0.45\textwidth]{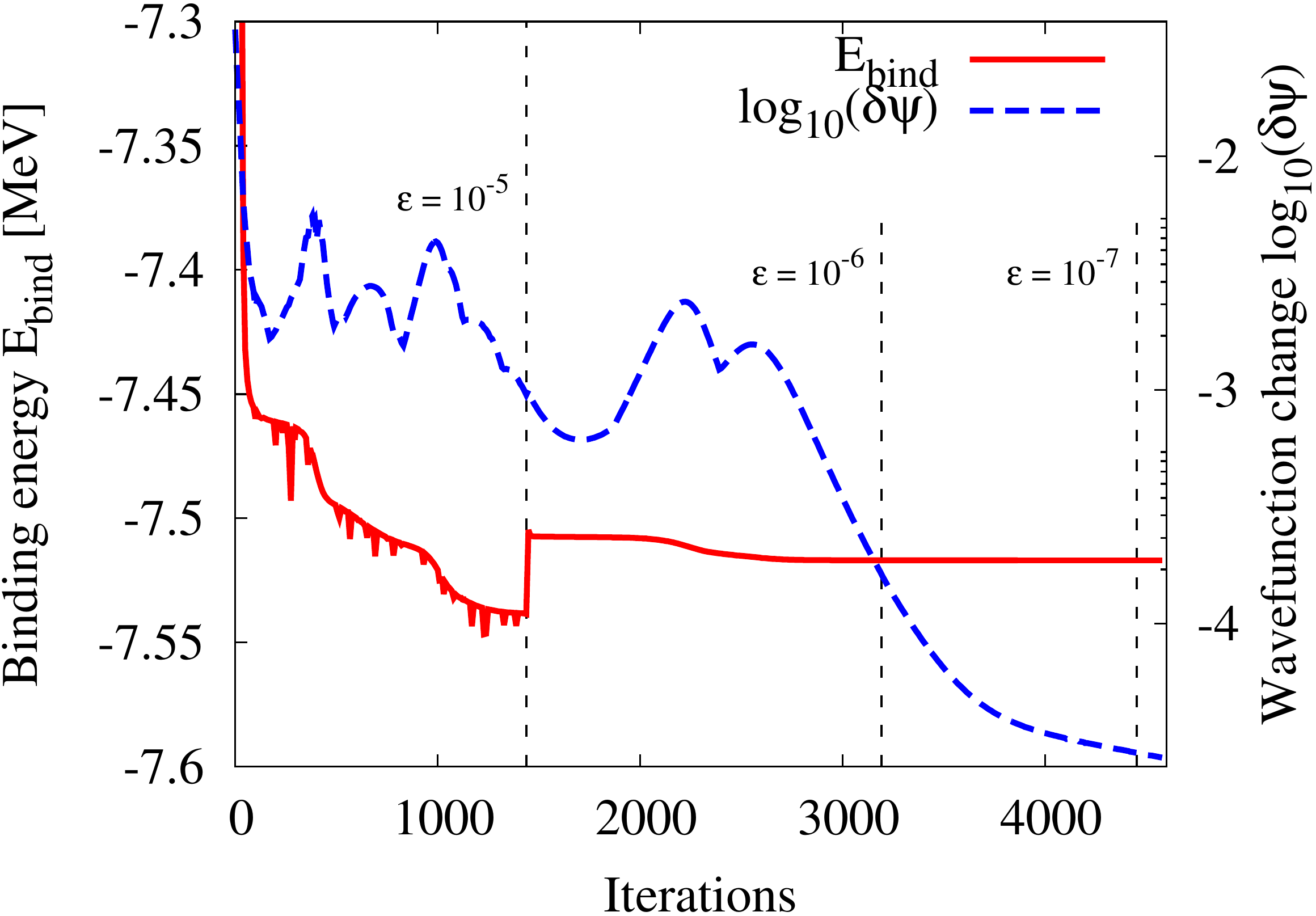}
\caption{Evolution of the maximum wavefunction change $\delta \psi$ and binding energy $E_\mathrm{bind}$ with iterations for the $^{238}\mathrm{U}$ nucleus.}
\label{U238_log}
\end{center}
\end{figure}

\begin{figure}
\begin{center}
\includegraphics[width = 0.22\textwidth]{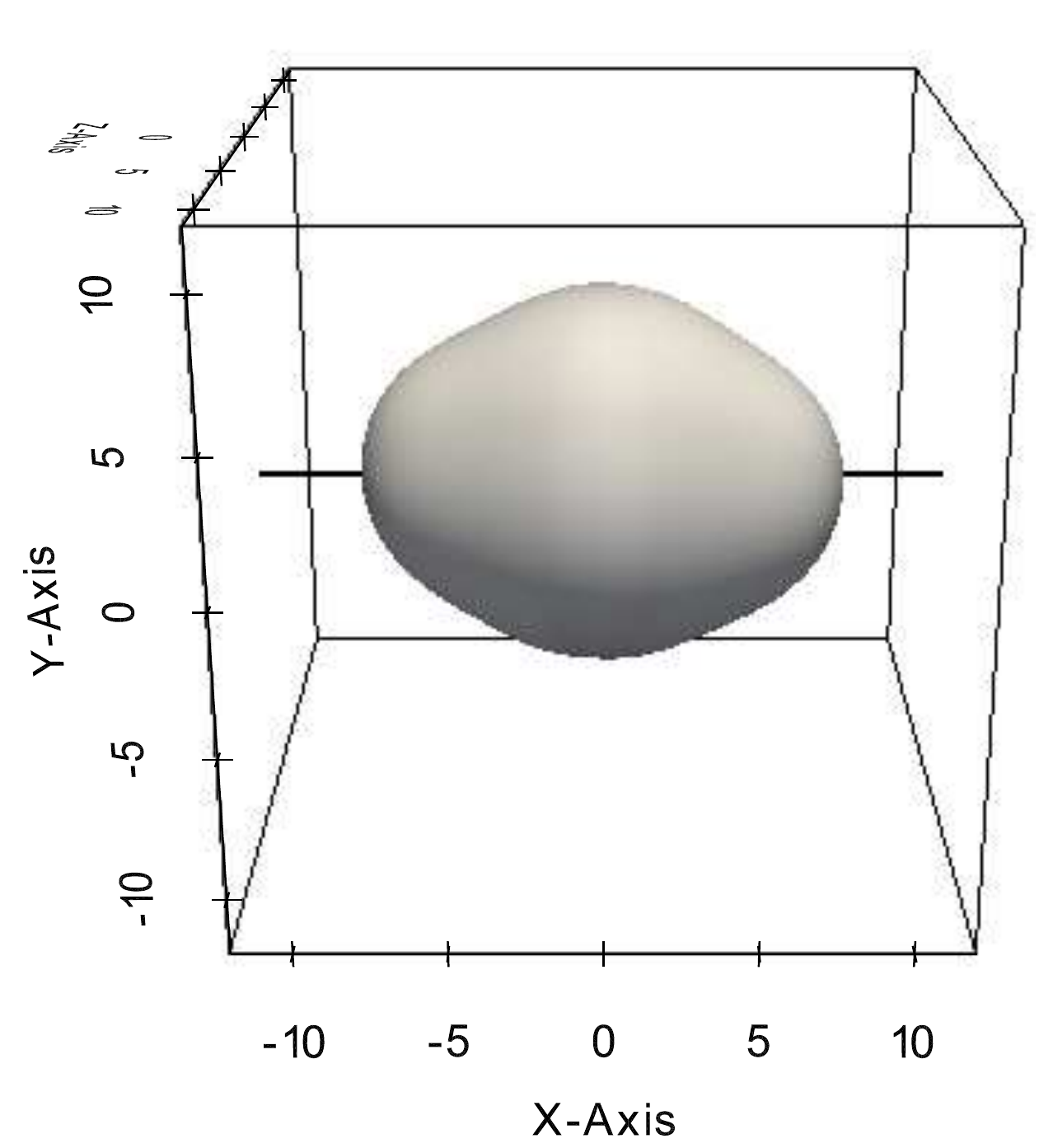}
\includegraphics[width = 0.22\textwidth]{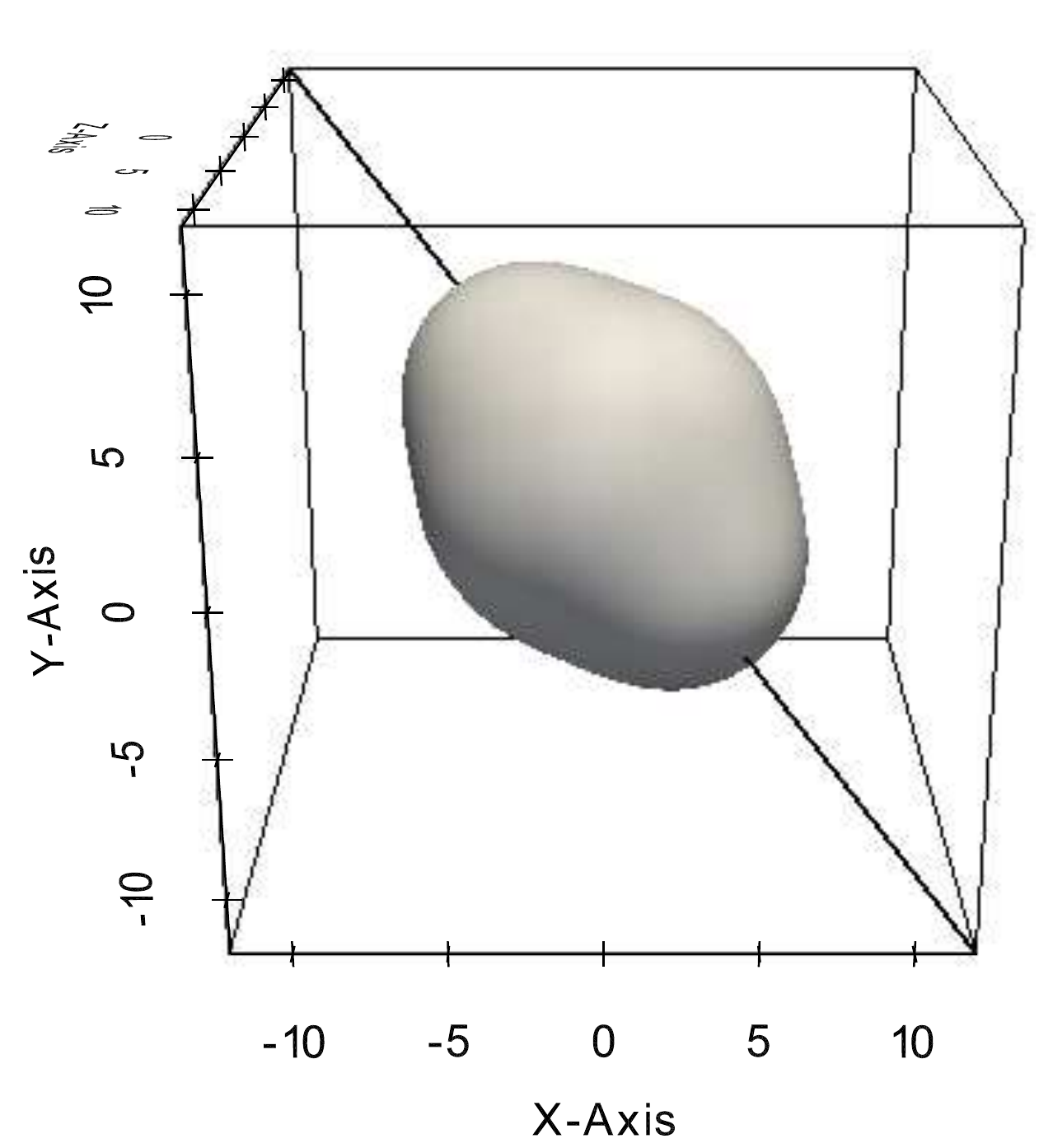}
\caption{Density iso-surface of the $^{238}\mathrm{U}$ nucleus for $\rho = 0.08 \: \mathrm{fm}^3$ for simulation (II) with $\epsilon = 10^{-7}$ and no Gaussian smoothing. The left figure shows the orientation of the nucleus in the simulation space at iteration 4000. The right figure shows the nucleus at iteration 19400. The black lines mark the symmetry axis.}
\label{U238_4000}
\end{center}
\end{figure}

\begin{table*}
\centering
\begin{tabularx}{\textwidth}{@{}  l c c c Y Y Y Y Y Y Y c c @{}  }                    
\hline
     & L [fm]&  sim. & resol. & E$_\mathrm{bind}$ &  E$_\mathrm{total}$ & E$_\mathrm{kin}$ & E$_0$ & E$_1$ & E$_2$ & E$_3$ & E$_4$ & E$_\mathrm{C}$ \\
\hline 
\hline 
Sky3D   & 24   &(I)   & 1 fm               & -7.521         & -1790.095 & 4493.687 & -19719.820 & 315.821 & 264.729 & 11991.680 & -90.798 & 954.640   \\
Sky3D   & 24   &(I)   & 0.5 fm            & \textbf{-7.521}         & -1790.056 & 4493.707 & -19720.940 & 315.855 & 264.758 & 11992.730 & -90.799 & 954.640   \\
\hline
M-SHF  & 200 &(I)$\star$   & $10^{-5} $ &-7.548           & -1796.439 & 4509.010 & -19808.196  & 316.196 & 264.335 & 12057.925 & -91.182 & 955.474  \\
M-SHF  & 200 &(I)   		  & $10^{-5} $  &-7.517                      & -1789.036 & 4493.844 & -19716.954  & 315.797 & 264.682 & 11989.662 & -90.682 & 954.613  \\
M-SHF  & 200 &(I)   		  & $10^{-7} $  &\textbf{-7.517}         & -1789.037 & 4493.815 & -19716.783  & 315.794 & 264.672 & 11989.534 & -90.681 & 954.612  \\
\hline
\hline
Sky3D   & 24   &(II)   & 1 fm               & -10.284                & -2447.609 & 4524.698 & -19932.520 & 322.707 & 267.612 & 12160.020 & -92.395 & 302.305  \\ 
Sky3D   & 24   &(II)   & 0.5 fm            & \textbf{-10.284}                & -2447.561 & 4524.779 & -19933.900 & 322.747 & 267.675 & 12161.240 & -92.414 & 302.315  \\
\hline
M-SHF  & 24 &(II)$\star$   & $10^{-5} $  &-10.318                      & -2455.747 & 4538.262 & -19997.608  & 322.283 & 267.574 & 12207.267 & -94.137 & 300.612 \\
M-SHF  & 24 &(II)               & $10^{-5} $  &-10.282                       & -2448.297 & 4523.168 & -19918.433  & 322.321 & 267.700 & 12149.303 & -92.444 & 300.090 \\
M-SHF  & 24 &(II)               & $10^{-7} $  & \textbf{-10.288}         & -2448.444 & 4523.097 & -19917.579  & 322.290 & 267.709 & 12148.620 & -92.475 & 299.893 \\
\hline  
\end{tabularx}
\caption{Parameters and energies for Sky3D and M-SHF simulations of the $^{238}$U nucleus. Table parameters are the same as in table \ref{table_O16}. The experimental binding energy is E$_\mathrm{bind} \sim 7.570\:$MeV \cite{Audi03}. }
\label{table_U238}
\end{table*}
Finally, we discuss the $^{238}\mathrm{U}$ nucleus. The density profiles for the initial and final states in simulation (I) are shown in Fig.\ref{U238}(a) for $-20\:\mathrm{fm} \leq x,y,z \leq 20\:$fm. As before, we start with harmonic oscillator states. The initial density distribution of $^{238}\mathrm{U}$ is again squeezed in the z-direction while    
the final nucleus is elongated along the y-axis. The many required shape changes result in a longer convergence time as will be discussed at the end of this section. Results for different simulations with Sky3D and M-SHF are given in table \ref{table_U238}. As previously mentioned, for Sky3d, we perform simulation with $\Delta x = 1\:$fm and $L=24\:$fm. A smaller value of $\Delta x$ and larger simulation space results in small changes in the energy contributions. As can be found in table \ref{table_U238}, final truncation thresholds of $\epsilon = 10^{-5}$ and $\epsilon = 10^{-7}$ in M-SHF do not lead to large differences in the energies. Gaussian smoothing, on the other hand, impacts all energy terms and results in $| \Delta E_\mathrm{total} | \sim 7.403\:$MeV for setup (I). This difference is larger than for $^{208}\mathrm{Pb}$. Figure \ref{U238}(b) shows a comparison between the x-profiles of $\rho$, $\tau$ and $\Delta \rho$ for the converged state with $\epsilon = 10^{-7}$ and $\epsilon = 10^{-5}$ whereas the latter contains smoothing. Small differences are present, especially in $\Delta \rho$. For better visualization we zoom in again and plot $\rho$, $\tau$ and $\Delta \rho$ for $0\:\mathrm{fm} \leq x \leq 8\:$fm, $0.145\:\mathrm{fm} \leq y \leq 0.18\:$fm and $-0.025\:\mathrm{fm} \leq y \leq 0.025\:$fm for $\rho$, $\tau$ and $\Delta \rho$, respectively in Fig. \ref{U238_sc_2}. We can now see that the differences between simulations with and without smoothing are more pronounced than for $^{208}\mathrm{Pb}$ and we assume that this leads to the larger value of $| \Delta E_\mathrm{total}|$. \\
Since its initial shape is very different from the final configuration, $^{238}\mathrm{U}$ has to undergo many shape changes during the iterations. On eight CPU nodes this results in simulations times on the order of days for about 4500 iterations with a final truncation threshold of $\epsilon = 10^{-7}$. The evolution of $\log(\delta \psi)$ and $E_\mathrm{bind}$ are shown in Fig. \ref{U238_log}.
Unlike the $^{208}\mathrm{Pb}$ simulation, $\log(\delta \psi)$ has now many local maxima and minima. These correspond to shape changes of the nucleus until the lowest energy state is found. The evolution of the binding energy follows a gradual decrease with some small fluctuations. The latter seem to be a result of the interplay of Gaussian smoothing with the evolving wave functions. As soon as smoothing is removed around iteration $1500$, the fluctuations in the binding energy disappear and the latter jumps to higher values. The experimental value for $^{238}\mathrm{U}$ is $E_\mathrm{exp} = -7.570 \: \mathrm{MeV}$ and thereby in good agreement with the simulations. \\
\begin{figure}
\begin{center}
\includegraphics[width = 0.45\textwidth]{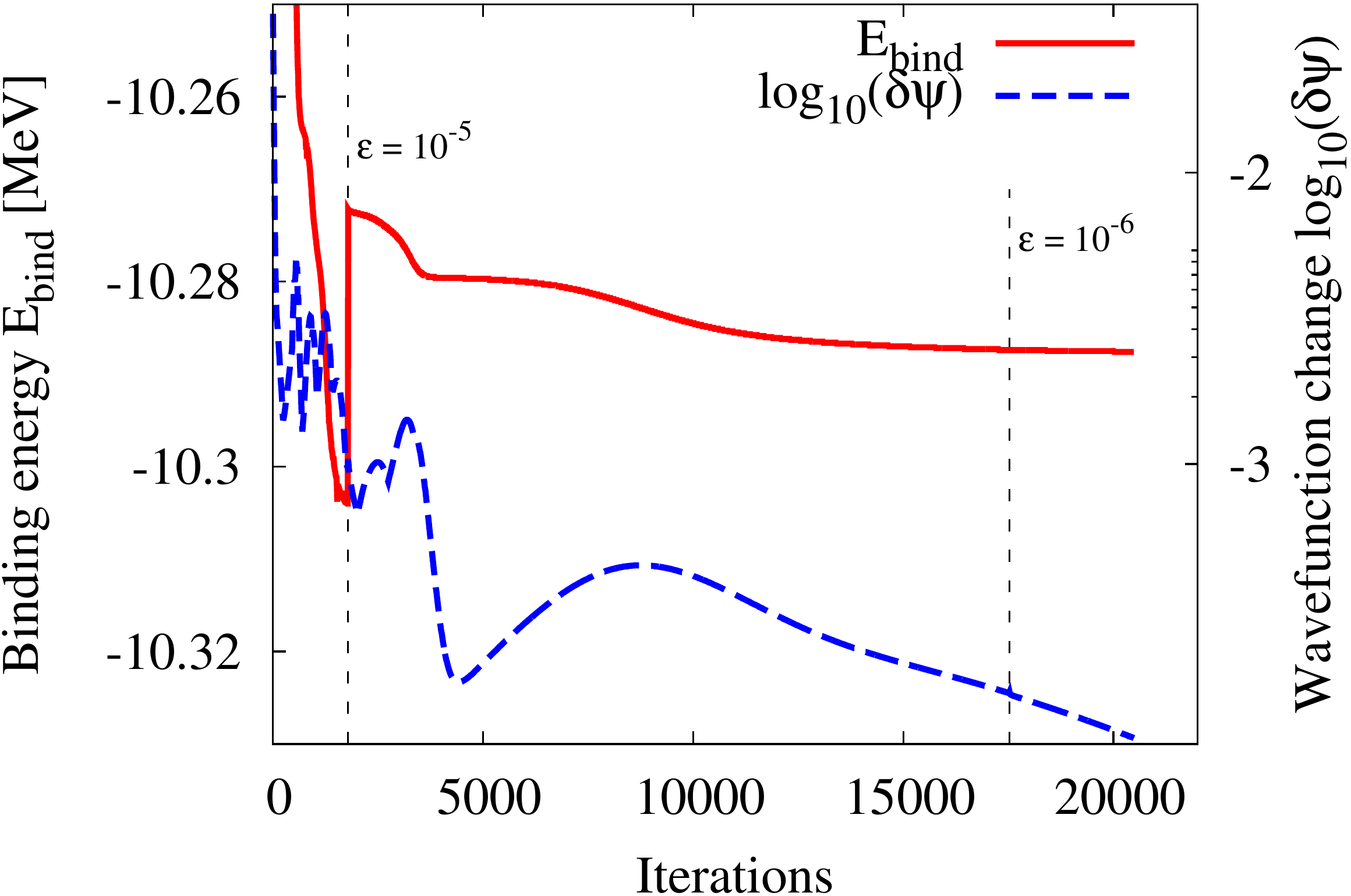}
\caption{Evolution of the maximum wavefunction change $\delta \psi$ and binding energy $E_\mathrm{bind}$ with iterations for the $^{238}\mathrm{U}$ nucleus for simulation setup (II), $\epsilon = 10^{-7}$ and no Gaussian smoothing.}
\label{U238_pbc_log}
\end{center}
\end{figure}
As before, simulations of type (II) are initialized in the same way as simulations (I). For $\epsilon = 10^{-7}$ and no Gaussian smoothing, both calculations setups evolve similarly up to iteration $\sim 4400$. At this point, as is shown in Fig.\ref{U238_pbc_log}, $\delta \psi$ of setup (II) stops to decrease and increases again. It reaches a maximum around iteration 8800 and then decreases slowly until iteration 21000 where we stop the simulation. At this point, the configuration has not yet reached convergence according to our criteria, however, $\delta \psi$ is very small, around $10^{-4}$ and is unlikely to grow again. Upon examining the cause of the increase in $\delta \psi$ we find a change in the orientation of the $^{238}\mathrm{U}$ nucleus. Figure \ref{U238_4000} shows its density iso-surface corresponding to $\rho = 0.08\: \mathrm{fm}^{-3}$ at iteration 4000 (left subfigure) and iteration 19400 (right subfigure). 
We can see that while the symmetry axis of the nucleus is initially parallel to the x-axis it changes to a diagonal of the simulation volume. This orientation seems to be energetically more favorable for periodic boundary conditions although the difference in binding energies between iteration 4000 and iteration 194000 is very small, around $\left| \Delta E_\mathrm{bind} \right| \sim 0.0106\:$MeV, as can be seen in Fig.\ref{U238_pbc_log}. It is encouraging that our code can find the energetically more favorable state. However, the required timescales and iteration numbers are very large. Simulations with $\epsilon = 10^{-5}$, with and without Gaussian smoothing, evolve similarly whereas they don't reach convergence within 21000 iterations. Furthermore, their symmetry axes are not perfect diagonals of the simulation volume as is the case for the calculation with $\epsilon = 10^{-7}$. The energies of all three calculations are given in table \ref{table_U238}. 
Interestingly, different to simulation (I), the $^{238}\mathrm{U}$ nucleus calculated with $\epsilon = 10^{-7}$ in the MADNESS simulation with periodic bc is more bound than for Sky3D. This is most likely due to the discussed rotation of the nucleus. Although the Sky3D simulation ran for about 30 000 iterations, the orientation of the $^{238}\mathrm{U}$ symmetry axis stayed parallel to the x-axis. As before, the nuclear configurations in simulations (II) are more bound than in simulations (I) due to the presence of jellium.\\
\newline
At this point, the findings regarding our M-SHF code can be summarized as follows: For small nuclei such as $^{16}\mathrm{O}$, the code is fast and can determine the ground state within a few iterations to high precision, for both, small and large simulation volumes. Although both converged nuclei, $^{16}\mathrm{O}$ and $^{208}\mathrm{Pb}$, are spherically symmetric making harmonic oscillator states are a good initial guess, the $^{208}\mathrm{Pb}$ simulations requires significantly more computational time. In MADNESS, the latter scales with truncation threshold $\epsilon$ which we adjust during the computation. Proper timing tests with a fixed truncation threshold will be done in the future and will give more detailed information on the scaling of the code. From our $^{238}\mathrm{U}$ simulations, we see that the M-SHF code can find the ground state of nuclear configuration through several shape and orientation changes and is in agreement with experimental binding energies and other SHF simulations. With that, we turn to the study of nuclear pasta. 
\section{Nuclear pasta from Molecular Dynamics}
\subsection{Without spin-orbit contributions}
\label{pasta}
Numerical Skyrme Hartree-Fock studies of nuclear pasta phases are usually performed by initializing single particle states as plane waves or random positioned Gaussians. Ideally, the nucleon wave functions then converge into the ground state configuration. However, as matter frustration allows for many different local energy minima and thereby pasta shapes, matter can easily become trapped in a quasi-ground state. To facilitate the search for the true ground state at a given density, proton fraction and temperature, restrictions can be placed on the symmetry and shape of the nuclear configuration \cite{Newton09}. This leads to a faster convergence of the latter and, by changing the symmetry assumptions, allows to scan through different pasta shapes. The ground state can then be identified as the one with the lowest total energy. The drawback of this approach is that the final shapes are somewhat predetermined by the specific assumptions and it might be difficult to explore new geometries.\\
Molecular dynamics simulations start out with randomly placed nucleons that are evolved over many iterations. The only restrictions for such methods are the imposed boundary conditions and simulation space dimensions. Different variations of the latter can be tested to ensure that the ground state is independent of the simulation space setup. Especially large molecular dynamics calculations with $\gtrsim 10^5$ nucleons and box lengths of $L \geq 100\:$fm minimize the effects of the simulation space geometry. The resulting pasta configurations can be traditional or novel shapes as found in \cite{Schneider13}. However, MD simulations often do not contain quantum mechanical features and their results should be cross-checked with self-consistent calculations such as Hartree-Fock. In this work, we use the converged pasta configuration of a simulation with the the Indiana University Molecular Dynamics code IUMD \cite{Horowitz04, Horowitz04b, Horowitz05, Horowitz08, Schneider13}  and explore its evolution as we iterate the single particle states with the M-SHF code.\\ 
\begin{figure*}
\centering{
\begin{minipage}{0.3\textwidth}
\includegraphics[width = 0.82\textwidth]{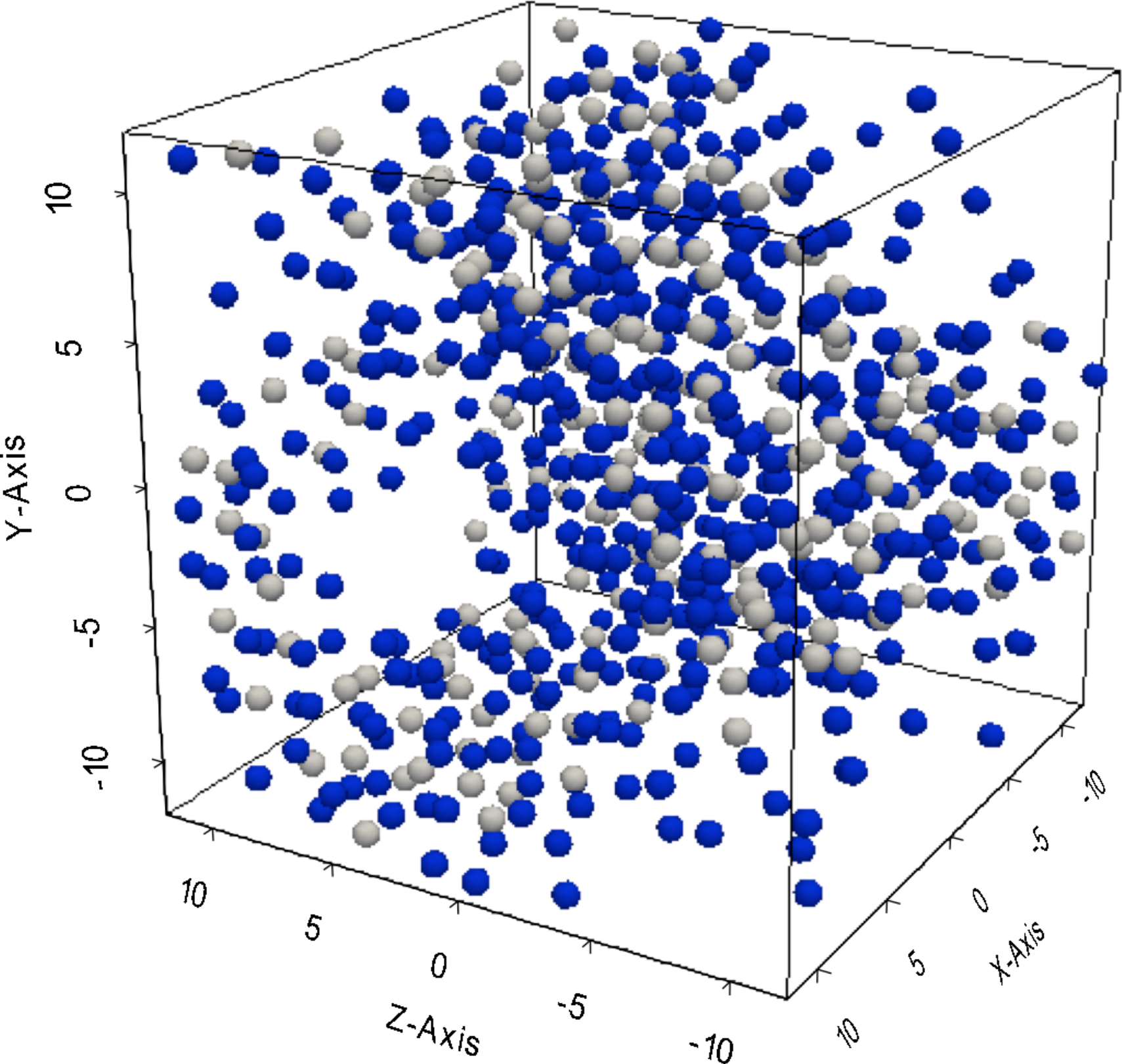}
\end{minipage}
\begin{minipage}{0.3\textwidth}
\includegraphics[width = 0.82\textwidth]{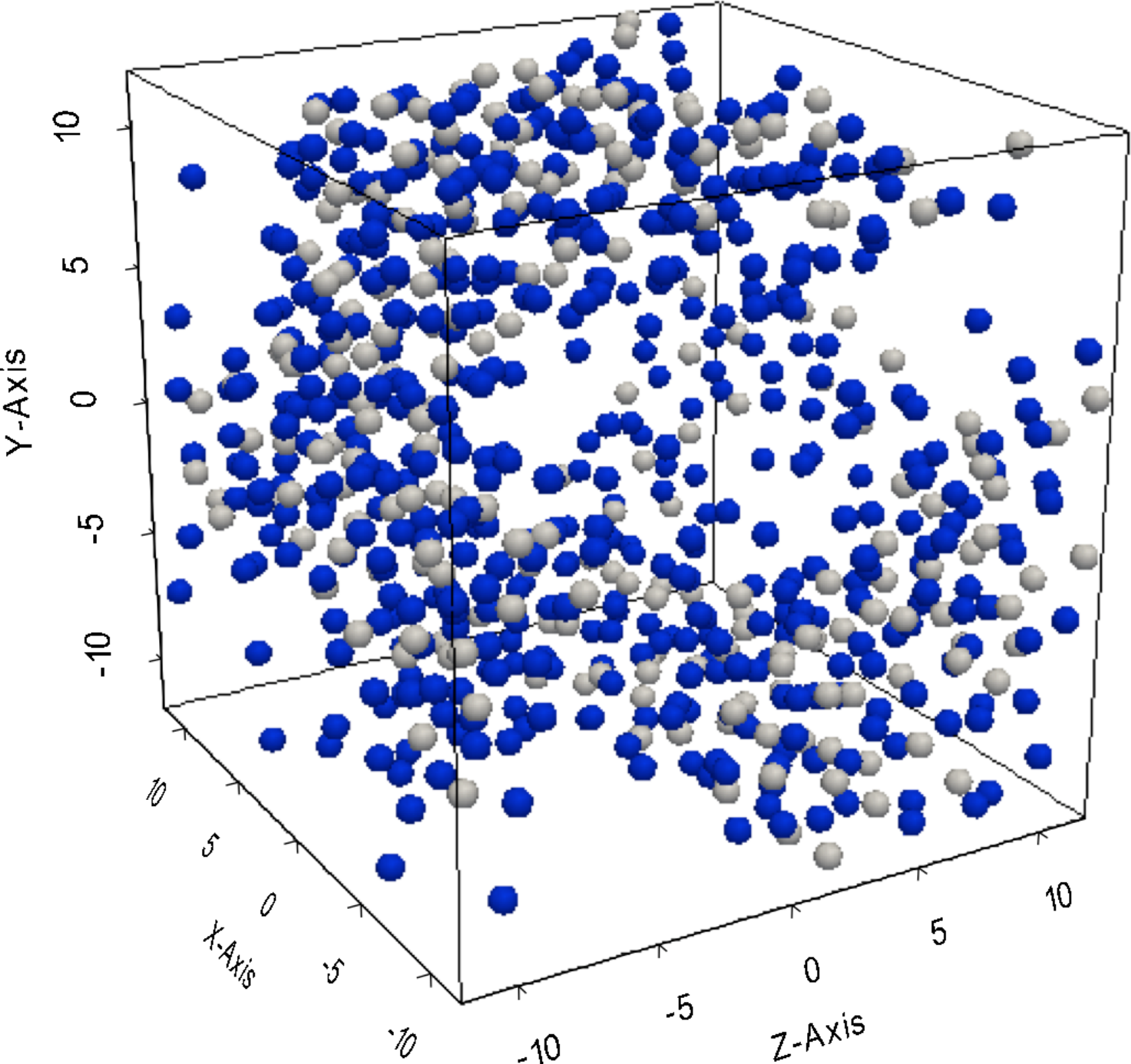}
\end{minipage}
\begin{minipage}{0.3\textwidth}
\includegraphics[width = 0.72\textwidth]{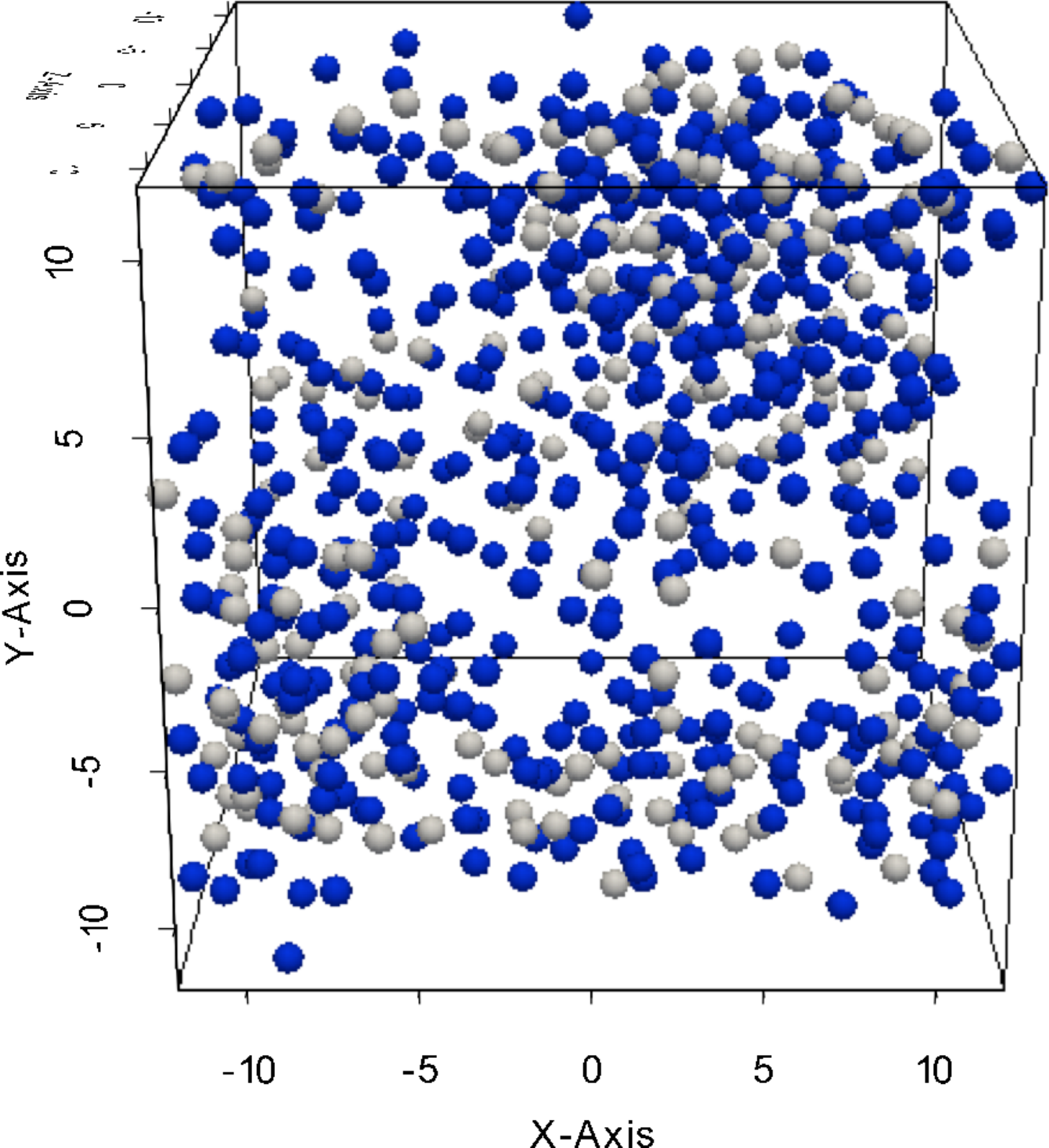}
\end{minipage}}
\caption{Positions of neutrons (blue) and protons (grey) in the converged IUMD pasta simulation. Subfigures show different orientations of the simulations space. Sphere sizes are for visualization only.}
\label{fig:md}
\end{figure*}
\begin{figure*}
\centering{
\begin{minipage}{0.32\textwidth}
\includegraphics[width = 0.99\textwidth]{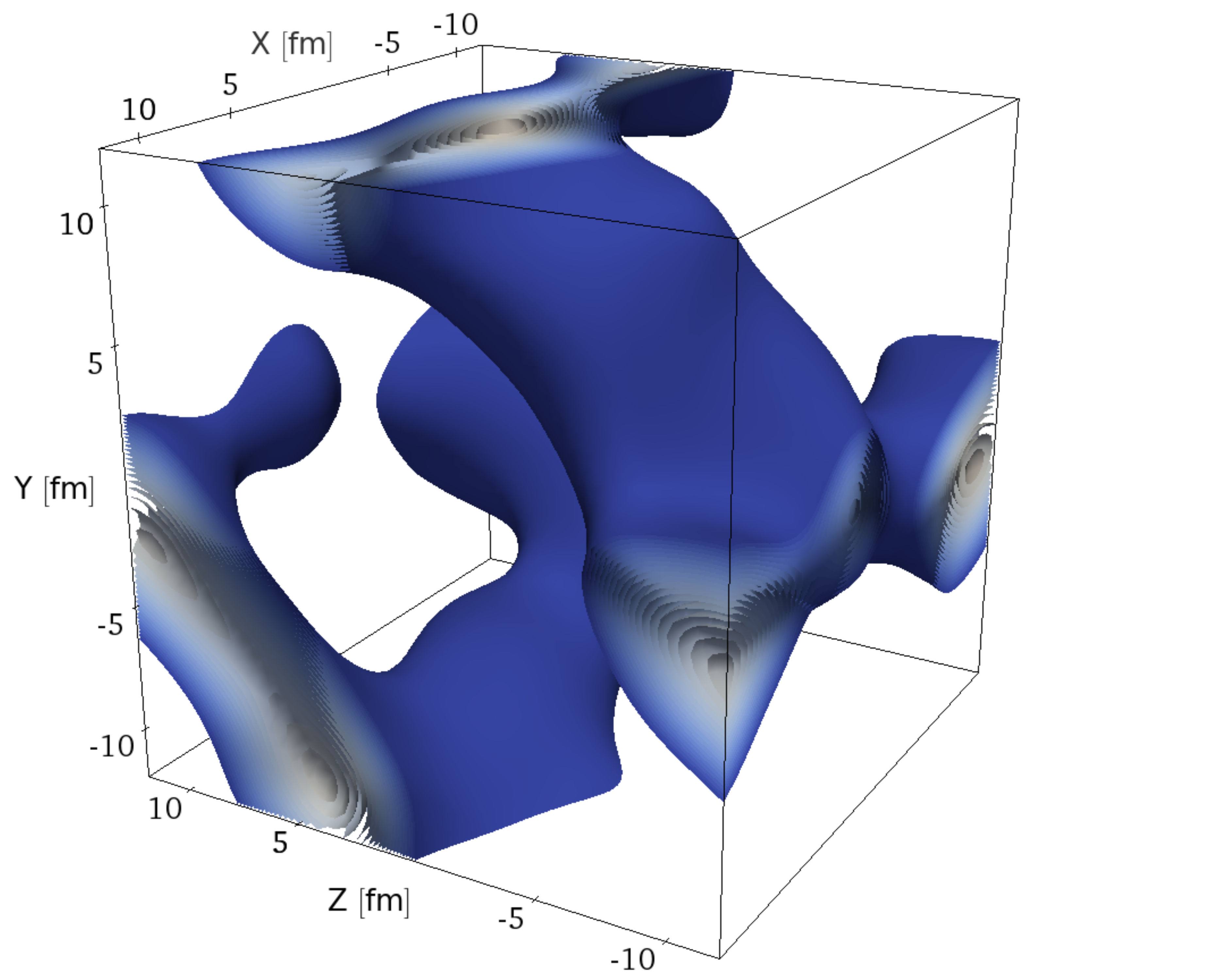}
\end{minipage}
\begin{minipage}{0.32\textwidth}
\includegraphics[width = 0.99\textwidth]{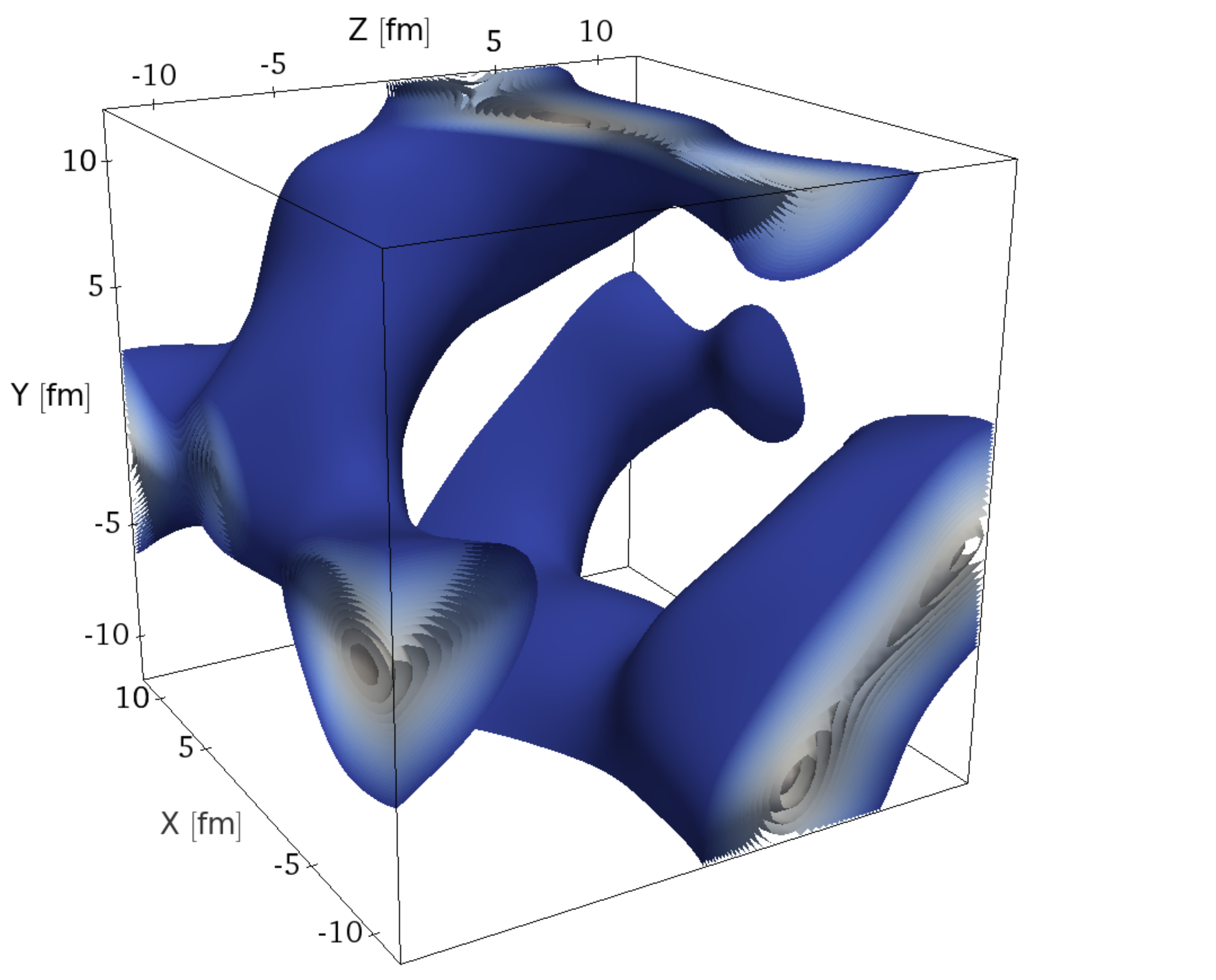}
\end{minipage}
\begin{minipage}{0.33\textwidth}
\includegraphics[width = 0.99\textwidth]{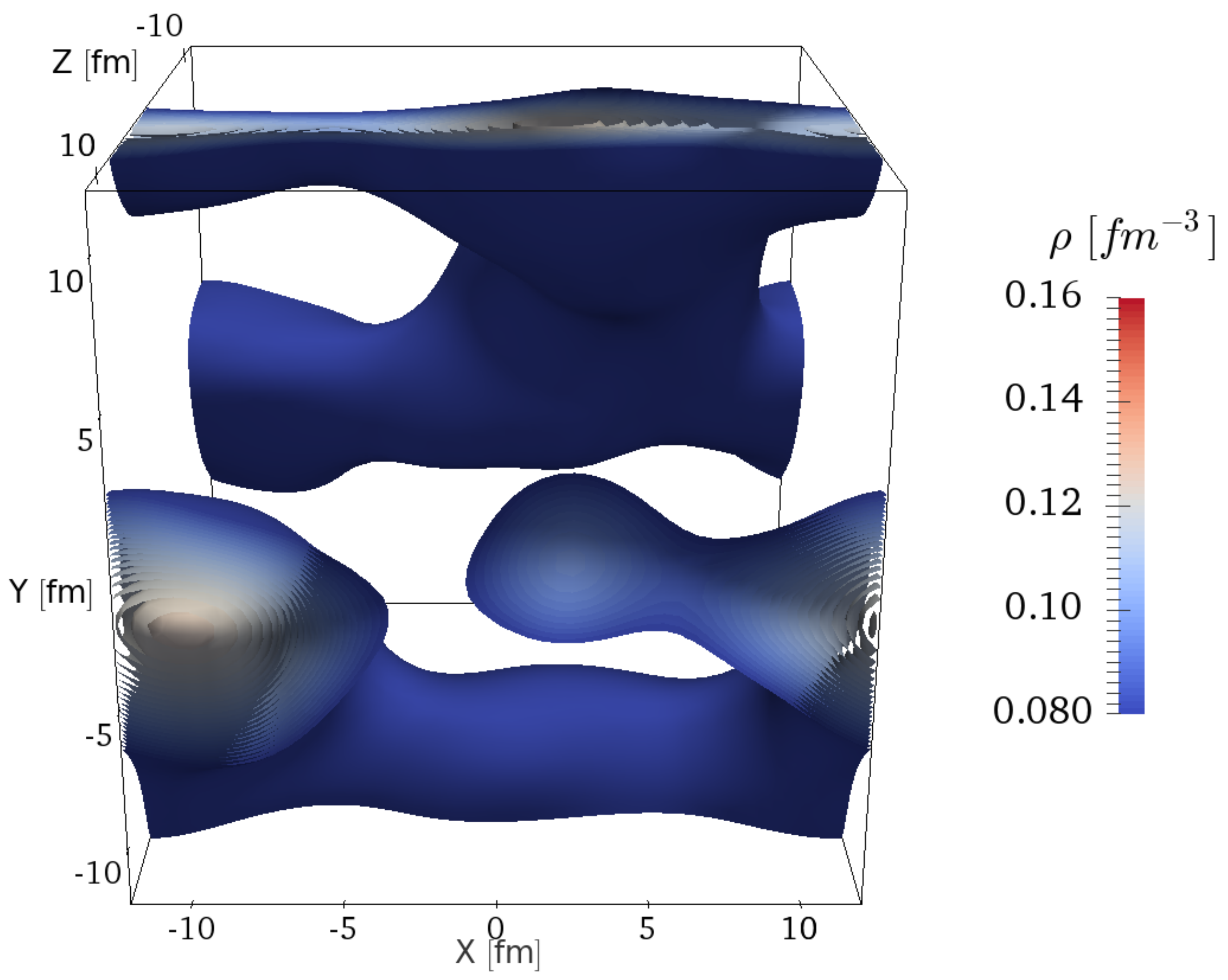}
\end{minipage}}
\caption{Iso-surfaces of initial total baryon number density $\rho (\vec{r}) $ of M-SHF. Nucleon wavefunctions are Gaussians folded around coordinates from IUMD shown in Fig.\ref{fig:md}. Subfigures correspond to orientations of the simulations space as in Fig.\ref{fig:md}. See text for details.}
\label{fig:pasta0}
\end{figure*}
\newline
Our starting point is the so-called waffle phase \cite{Schneider13} - an intermediate state between the lasagna (plate) and the spaghetti (rod) configurations. It consists of plates with a lattice of periodic holes whereas two neighboring plates are displaced by half of the lattice spacing. The IUMD calculation was performed using periodic boundary conditions with 490 neutron and 210 proton particles in a simulation box with length $L=24\:\mathrm{fm}$. The average density is $\rho = 0.05 \:\mathrm{fm}^{-3}$ with a proton fraction $Y_p = 0.3$. The converged nucleon positions are shown in Fig.\ref{fig:md} with blue (dark) spheres symbolizing neutrons and grey (light) spheres protons. Note that the temperature of the MD simulation was 1\:MeV while our M-SHF calculations are performed at zero temperature. Typical pasta calculations are performed either at supernova conditions - finite temperature and $Y_p \sim 0.2 - 0.4$ - or neutron star conditions with $T \sim 0$ and $Y_p \leq 0.1$. The configuration that we are describing, i.e. $T = 0$\:MeV and $Y_p = 0.3$, does therefore not exist in neutron stars or supernova environments. However, the aim of this paper is to explore the stability of the IUMD pasta configuration once quantum mechanical features are added and to provide a proof-of-principle study for future finite temperature M-SHF simulations that will apply neutron star or supernovae conditions.\\
For the initialization, each single particle state is given by a sum of 27 3D Gaussians with $\sigma = 3.0\:$fm. The Gaussian are centered at the nucleon coordinates from IUMD and their closest images due to the reflective boundary conditions. Figure \ref{fig:pasta0} shows the resulting density iso-surfaces for $0.08 \: \mathrm{fm}^{-3} \leq \rho \leq 0.16 \: \mathrm{fm}^{-3}$ and different orientations of the simulation space. We can identify two plates, each with one hole. The latter are displaced relative to each other so that a hole is aligned with a denser region in the neighboring plate. The M-SHF iterations are performed as previously discussed for setup (II). Here, we do not include either the spin-orbit potential $U_{so}$
\begin{figure}
\begin{center}
\includegraphics[width = 0.45\textwidth]{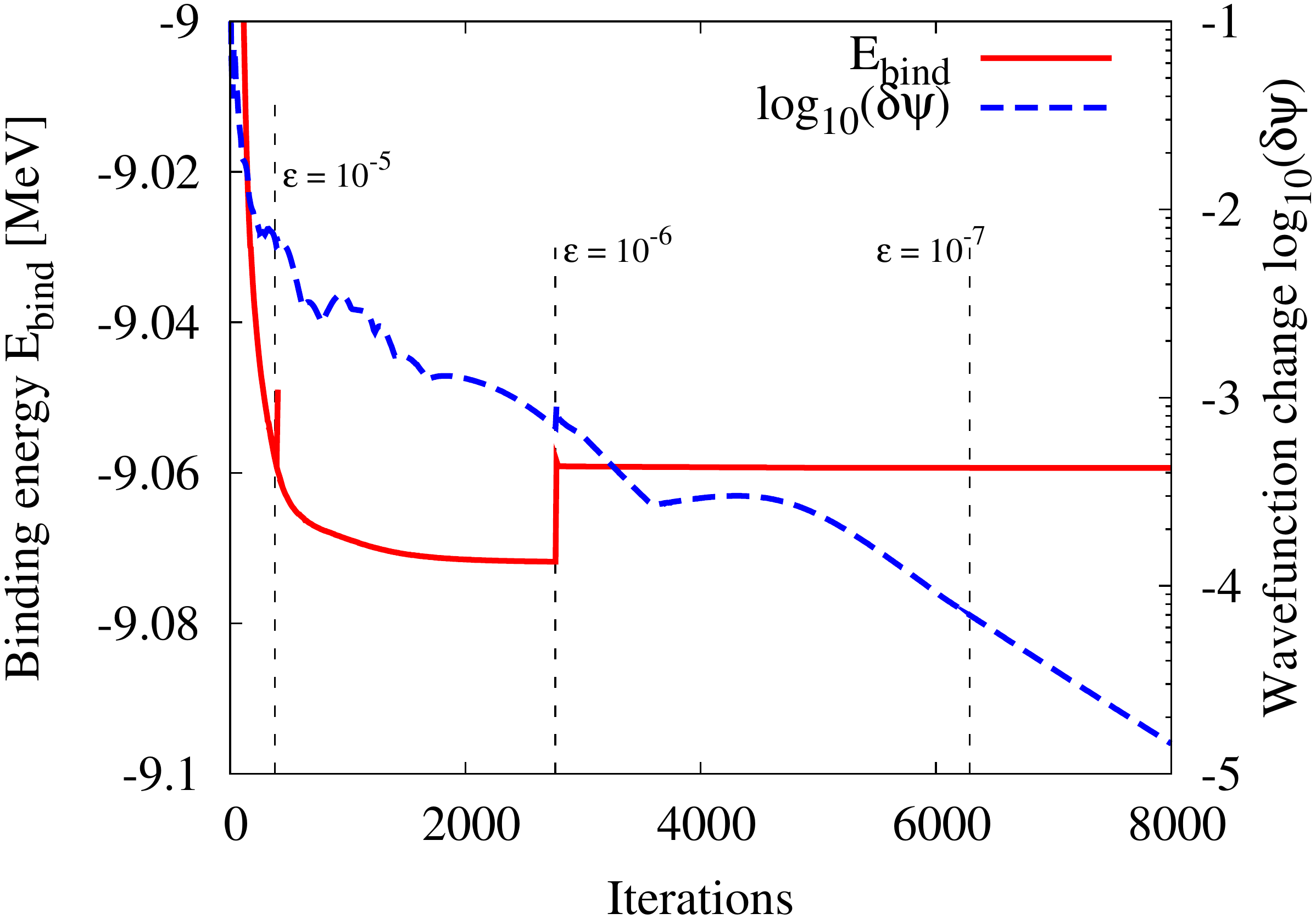}
\caption{Evolution of the maximum wavefunction change $\delta \psi$ and binding energy $E_\mathrm{bind}$ with iterations for the MD M-SHF pasta simulation}
\label{MD_log}
\end{center}
\end{figure}
or the spin-orbit density $\vec{J}$ and will test their effects in the next section.\\
As for the nuclear ground states in the previous discussion, we plot the evolution of the maximum wave function change $\delta \psi$ and binding energy $E_\mathrm{bind}$ in Fig.\:\ref{MD_log} with vertical lines indicating the reduction of $\epsilon$. The simulation was stopped after $\sim 8000$ iterations when $\delta \psi \sim 10^{-5}$. It ran for about two weeks on 24-30 nodes. 
\begin{figure*}
\centering{
\begin{minipage}{0.32\textwidth}
\includegraphics[width = 0.99\textwidth]{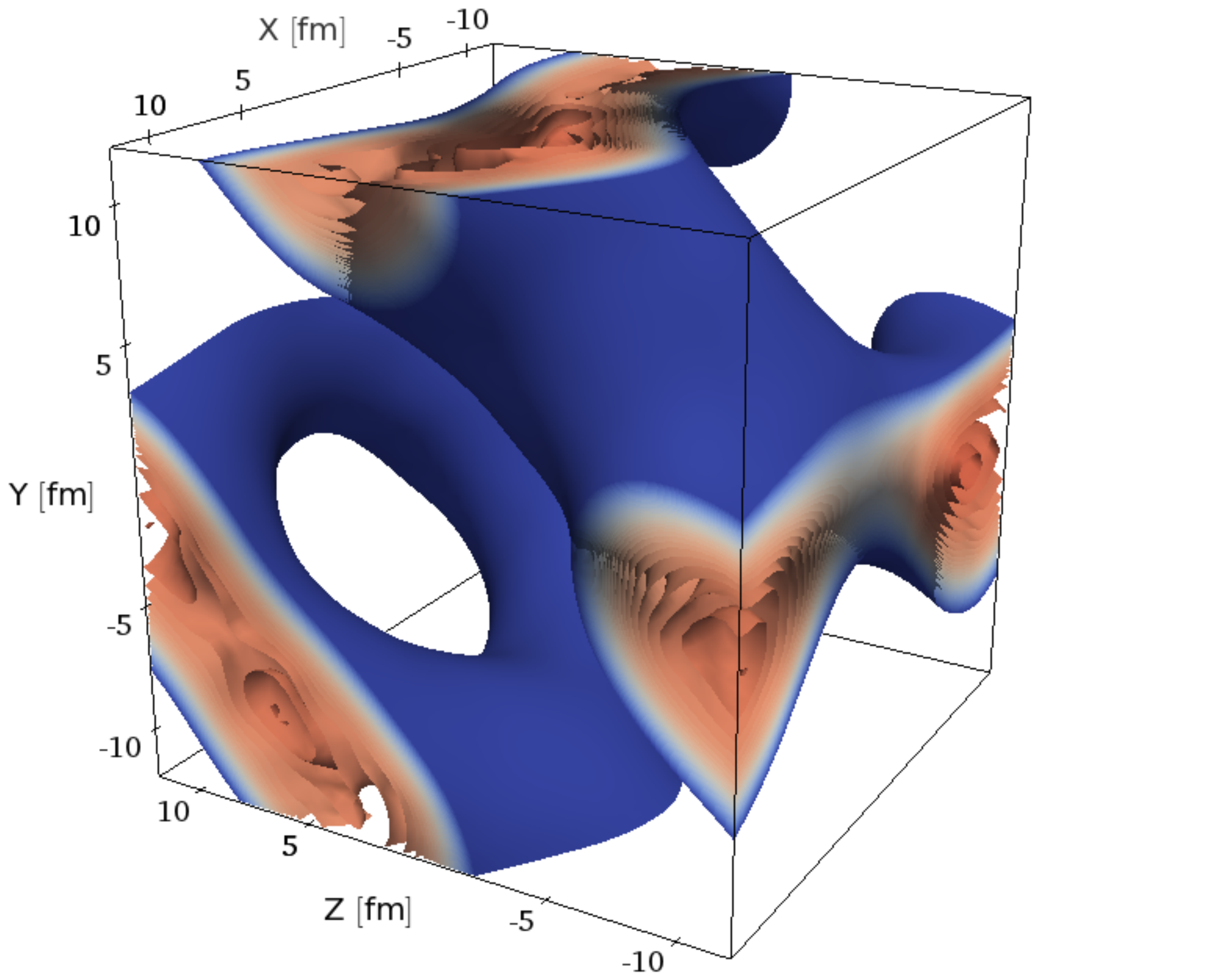}
\end{minipage}
\begin{minipage}{0.32\textwidth}
\includegraphics[width = 0.99\textwidth]{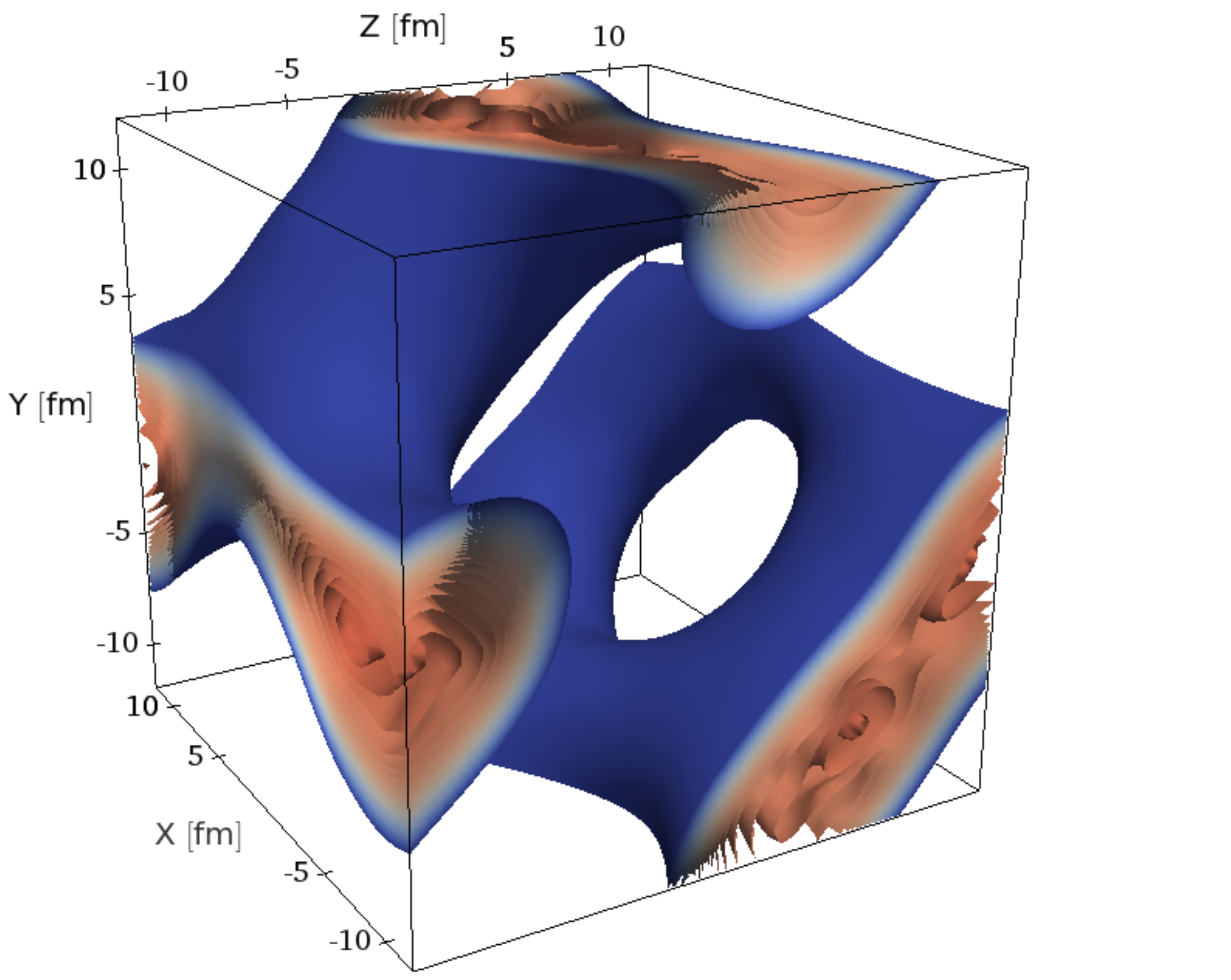}
\end{minipage}
\begin{minipage}{0.33\textwidth}
\includegraphics[width = 0.99\textwidth]{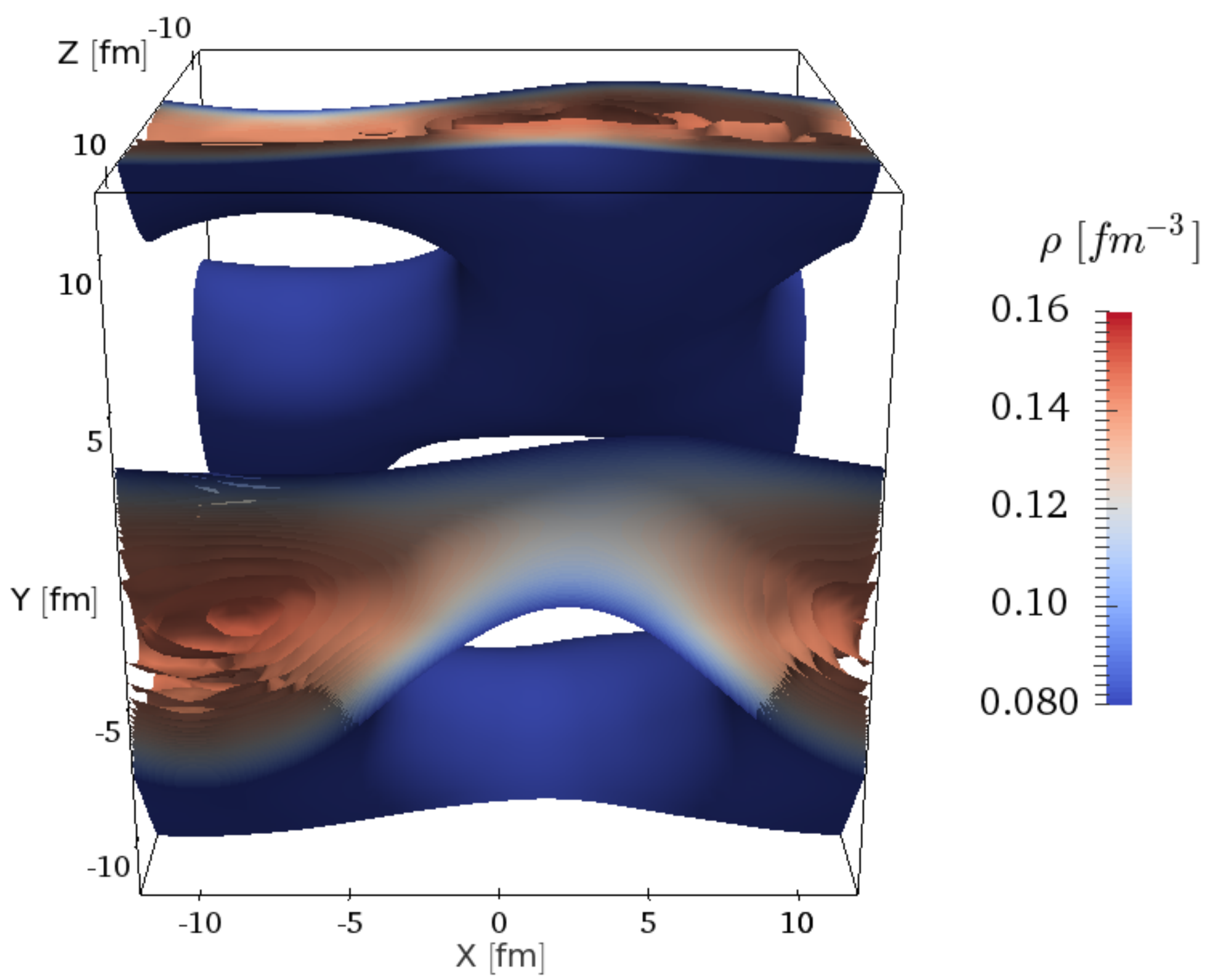}
\end{minipage}}
\caption{Iso-surfaces of the total baryon density $\rho (\vec{r})$ of the converged M-SHF simulation at iteration 8000. Subfigures are as in Fig.\ref{fig:pasta0}).}
\label{fig:pasta}
\end{figure*}
\begin{table*}
\centering
\begin{tabularx}{\textwidth}{@{}  l c c c Y Y Y Y Y Y Y c c @{}  }                    
\hline
     & L [fm]&  sim. & resol. & E$_\mathrm{bind}$ &  E$_\mathrm{total}$ & E$_\mathrm{kin}$ & E$_0$ & E$_1$ & E$_2$ & E$_3$ & E$_4$ & E$_\mathrm{C}$ \\
\hline 
M-SHF  & 24 &(II)   &  $10^{-7} $  &-9.059  & -6341.536 &  12997.724 & -51349.350  & 529.618 & 449.606 & 30803.440 & -  & 227.427 \\
Sky3D  & 24 &(II)    &  $1$ fm        &-9.083  & -6358.075 &  13019.705 & -51434.630  & 525.969 & 451.538 & 30853.410 & -  & 225.933 \\
\hline  
\end{tabularx}
\caption{Parameters and energies for the nuclear pasta simulation with M-SHF and Sky3D. The table setup is as in table \ref{table_O16}.}
\label{table_pasta}
\end{table*}
From the maxima and minima in $\delta \psi$, we see that the pasta shapes underwent some small changes in the beginning of the simulation for iterations $< 4600$. Since the reduction of the truncation threshold from $\epsilon = 10^{-4}$ to $\epsilon = 10^{-5}$ occurs very early in the simulation, we decided to apply Gaussian smoothing until the second reduction of $\epsilon$ to $10^{-6}$. This explains the jump in binding energy around iteration 2800. The binding energy then converges to a value of $E_\mathrm{bind} \sim - 9.059 \: \mathrm{MeV}$. The different energy components are given in table \ref{table_pasta} together with the results of the Sky3D pasta simulation. The latter also initialized the wavefunctions also via Gaussians folded around the MD nucleon coordinates. The Sky3D calculation converged after ca. 32000 iterations. The difference in total binding energies for Sky3D and M-SHF is $| \Delta E_\mathrm{total} | \sim 16.539\:$MeV which is about $\sim 2.60 \times 10^{-3} | E_\mathrm{total}| $. \\
\begin{figure}
\includegraphics[width = 0.45\textwidth]{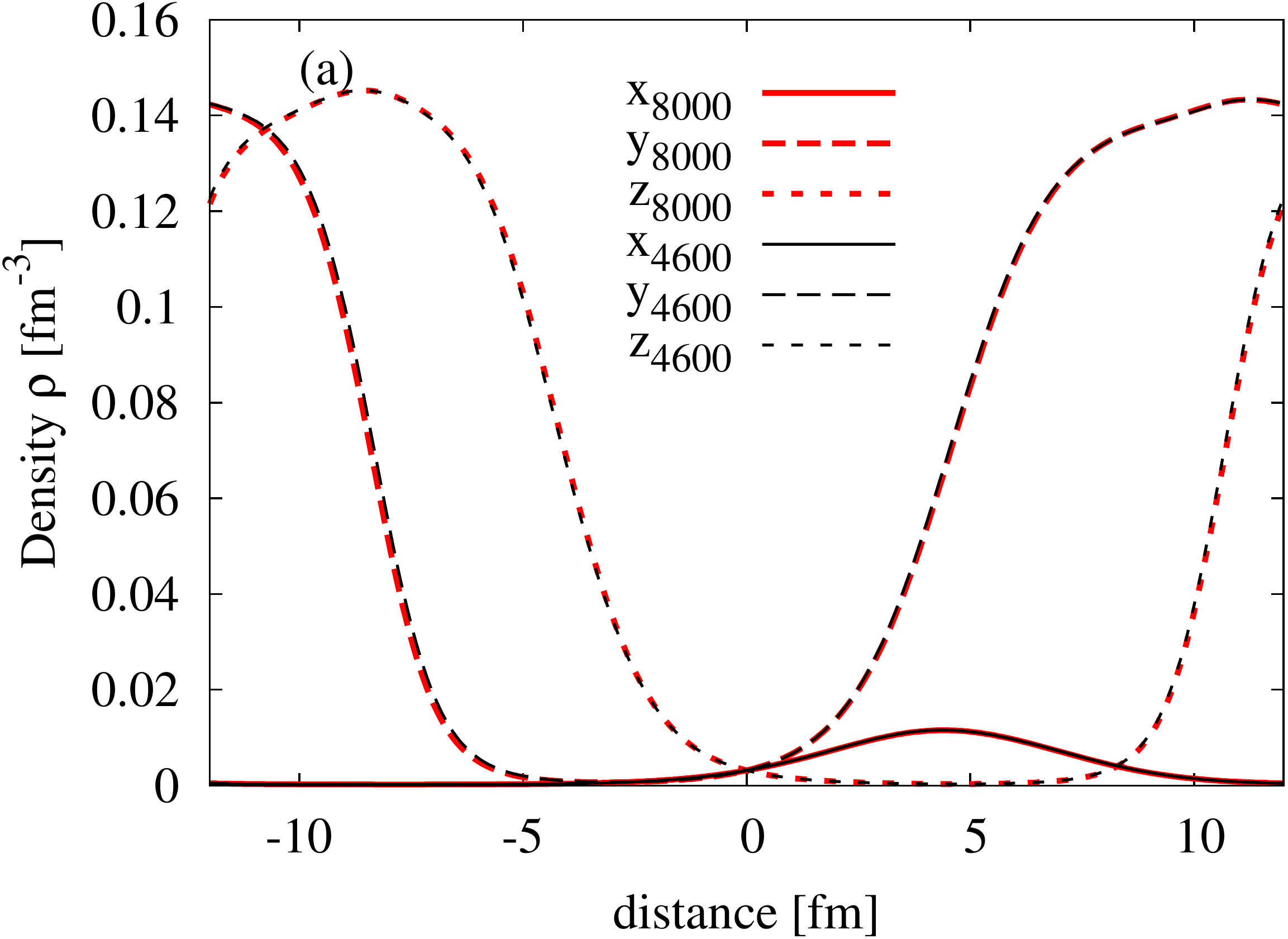}
\includegraphics[width = 0.45\textwidth]{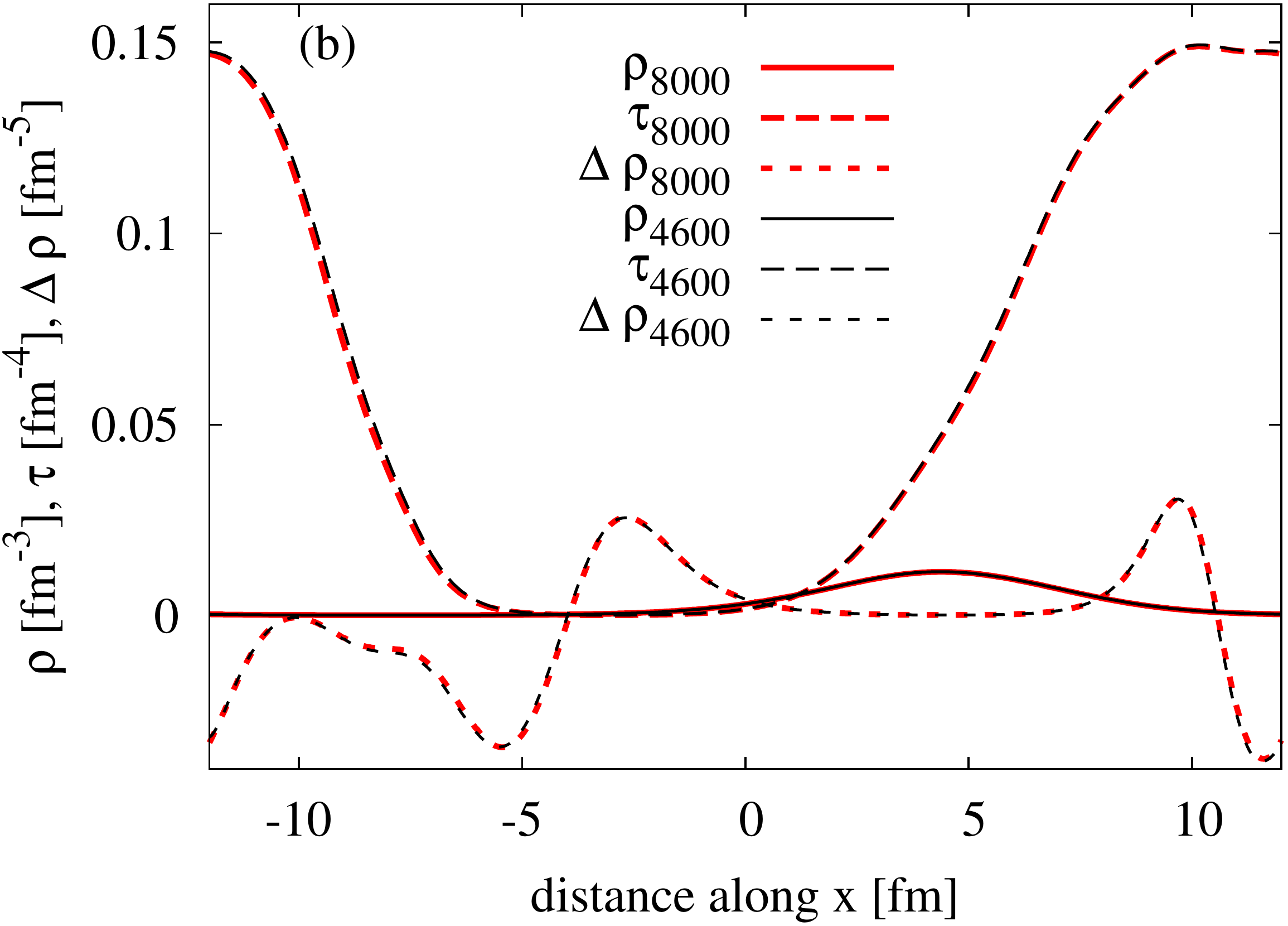}
\caption{(a) Number density profiles of the nuclear pasta configuration along the x, y, and z axis at iteration 4600 and in the final, i.e. converged, state at iteration 8000. (b) Comparison of the x-profiles of the total number density $\rho$, kinetic density $\tau$ and laplacian of the number density $\Delta \rho$ in the final, i.e. converged, state of the nuclear pasta and at iteration 4600.}  
\label{pic_pasta}
\end{figure}
Figure \ref{fig:pasta} shows the density iso-surfaces, again for $0.08 \: \mathrm{fm}^{-3} \leq \rho \leq 0.16 \: \mathrm{fm}^{-3}$ in the final M-SHF pasta configuration at iteration 8000. As before, subfigures correspond to different box orientations. The general shape of the waffle phase is very similar to Fig.\:\ref{fig:pasta0}. Main differences are the broadening of regions with $\rho \geq 0.08 \: \mathrm{fm}^{-3}$, while the holes in both plates are smaller. The variations of $\delta \psi$ correspond to these changes, whereas the smooth decrease of $\delta \psi$ for iterations $>4600$ implies that the ground state configuration at iteration 8000 is already reached at this point. Figure \ref{pic_pasta}(a) compares the number density profiles of the pasta configuration at iteration 4600 and 8000. The differences are very small which also applies to the kinetic densities and the laplacian of the densities along the x-axis, as shown in Fig.\ref{pic_pasta}(b). For future studies we should therefore consider to either modify the convergence criteria, consider a configuration to be stable at an earlier point in the simulation, or change the iteration procedure so that the convergence criteria are met faster once the pasta shape does not undergo significant changes.\\ 
\subsection{With spin-orbit contributions}
\label{pasta}
\begin{figure}
\begin{center}
\includegraphics[width = 0.45\textwidth]{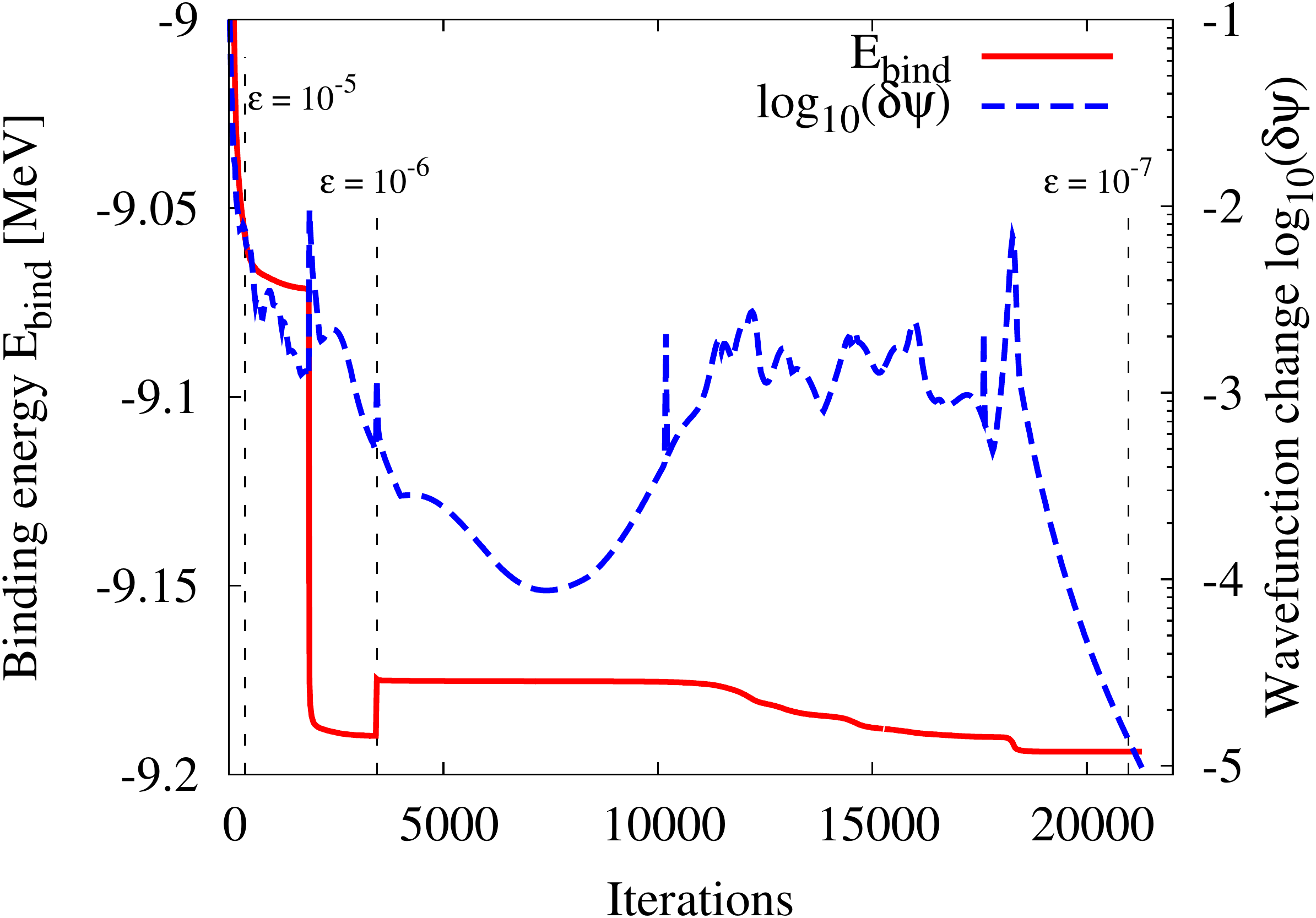}
\caption{Evolution of the maximum wavefunction change $\delta \psi$ and binding energy $E_\mathrm{bind}$ with iterations for the MD M-SHF pasta simulation}
\label{MD_log_so}
\end{center}
\end{figure}

Although, the spin-orbit contribution is not expected to be a large part of the total nuclear binding energy, it is important for the reproduction of energy levels and magic numbers. It has also been found that the spin-orbit potential can impact the deformation of nuclei \cite{Takahara11}, however, the effect on nuclear pasta phases has not been fully explored yet.\\
In this section we study the impact of the spin-orbit potential on the waffle pasta phase. We perform the calculations with Sky3D and M-SHF. For the latter, the initial configuration is taken from the previous section at iteration $\sim 2000$ when $\delta \psi \sim 10^{-3}$ (see Fig.\ref{MD_log}). For Sky3D, the starting point is the converged MD state. Here, as in the previous study, Gaussians are folded around the nucleon coordinates and evolved. However, now the nuclear potential contains the spin-orbit terms from the very beginning. For M-SHF, Fig.\ref{MD_log_so} shows the evolution of $\delta \psi$ and $E_\mathrm{bind}$ as functions of iteration number. Vertical dashed lines indicate the reduction of the truncation threshold. Note that the reduction of $\epsilon$ from $10^{-6}$ to $10^{-7}$ happens quite late in the simulation. Due to speed we evolved the configuration to lower $\delta \psi$ with $10^{-6}$ and reduced the truncation threshold only at the very end of the simulation. However, the switch from $\epsilon = 10^{-6}$ to $10^{-7}$ and consequent evolution over ca. 400 iterations did not cause any changes in the pasta configuration or its convergence.\\
A large jump can be seen in $\delta \psi$ as well as a step-like increase in $| E_\mathrm{bind} |$ as soon as we add the spin-orbit terms. The second jump around iteration 3500 when we reduce the truncation threshold from $\epsilon = 10^{-5}$ to $10^{-6}$ is due to the removal of Gaussian blurring. The simulation converges until iteration $\sim 7500$ where $\delta \psi$ starts to increase again. The different minima and maxima between iterations 10000 - 18000 indicate possible shape changes. Eventually, $\delta \psi$ starts to continuously decrease and reaches convergence with $\delta \psi \sim 10^{-5}$ for $\epsilon = 10^{-7}$. In total, the simulation requires about 21000 iteration steps whereas the Sky3D calculation converges already after iteration 6000. It is not clear whether adding the spin-orbit potential from the very beginning would also lead to a quicker convergence for M-SHF. This has to be explored in the future. \\
\begin{table*}
\centering
\begin{tabularx}{\textwidth}{@{}  l c c c Y Y Y Y Y Y Y c c @{}  }                    
\hline
     & L [fm]&  sim. & resol. & E$_\mathrm{bind}$ &  E$_\mathrm{total}$ & E$_\mathrm{kin}$ & E$_0$ & E$_1$ & E$_2$ & E$_3$ & E$_4$ & E$_\mathrm{C}$ \\
\hline 
M-SHF  & 24 &(II)   &  $10^{-7} $  &-9.194  & -6435.792 &  13327.675 & -52750.085  & 550.097 & 486.129 & 31884.606 & -168.597  & 234.383 \\
Sky3D  & 24 &(II)    &  $1$ fm        &-9.201  & -6440.795 & 13329.782 & -52767.190  & 549.622 & 487.920 & 31897.320 &  -171.994 & 233.754 \\
\hline  
\end{tabularx}
\caption{Parameters and energies for the nuclear pasta simulation with M-SHF. The table setup is as in table \ref{table_O16}.}
\label{table_pasta_so}
\end{table*}
\newline
Table \ref{table_pasta_so} compares the final energies of the two simulations. The energy contributions are similar. As before, we find that the absolute of the binding energy is smaller for M-SHF than for Sky3D. The difference is about $| \Delta E_\mathrm{total} | \sim 5\:$MeV which is only $\sim 7.77 \times 10^{-4} | E_\mathrm{total}|$. With spin-orbit, the waffle phase is more bound, by about $\sim 82.72\:$MeV and $\sim 94.26\:$MeV for Sky3D and M-SHF, respectively, which is $\sim 1.3 - 1.5 \%$ of the total energy.\\
\begin{figure*}
\centering{
\begin{minipage}{0.32\textwidth}
\includegraphics[width = 0.92\textwidth]{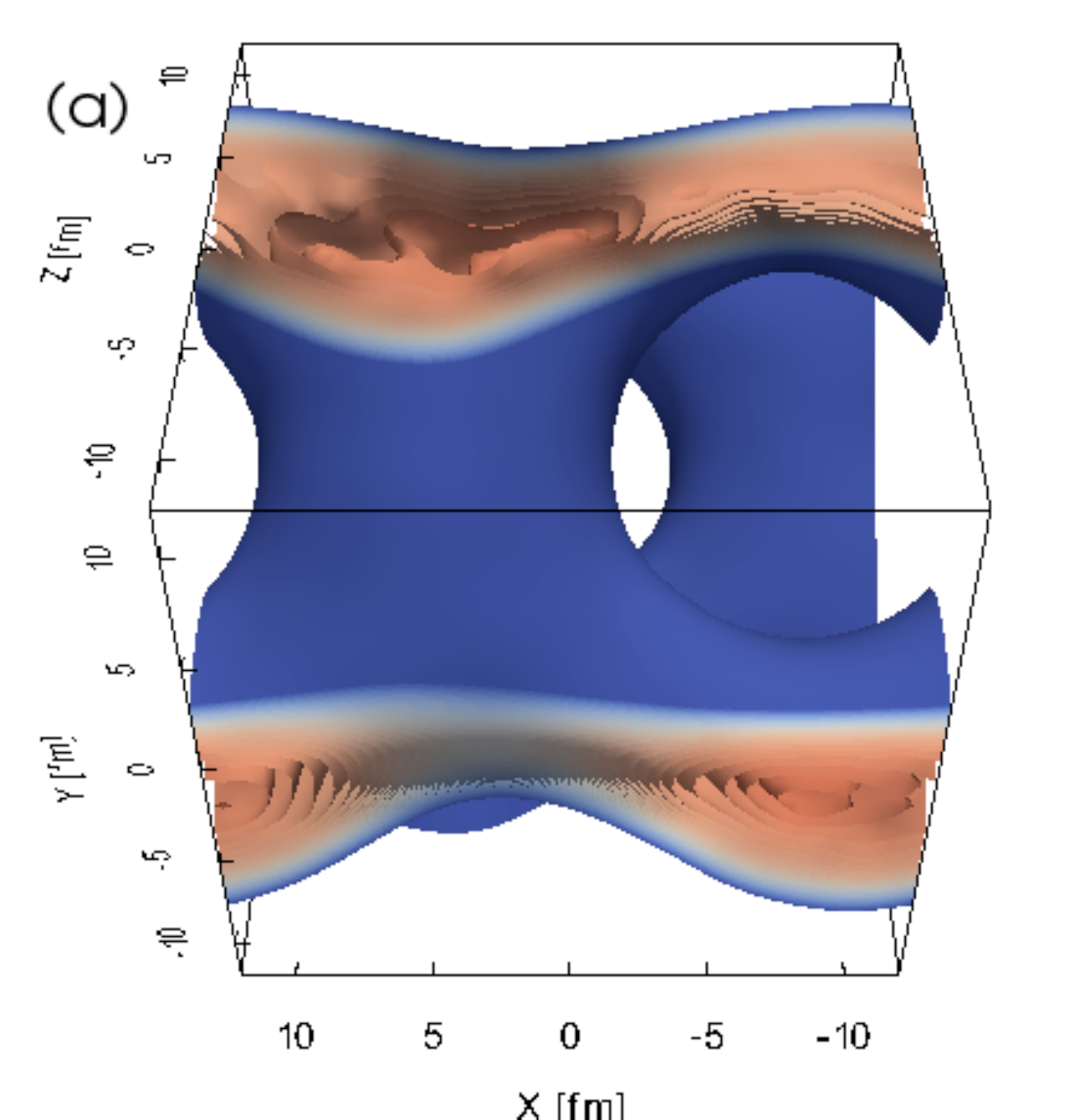}
\end{minipage}
\begin{minipage}{0.32\textwidth}
\includegraphics[width = 0.92\textwidth]{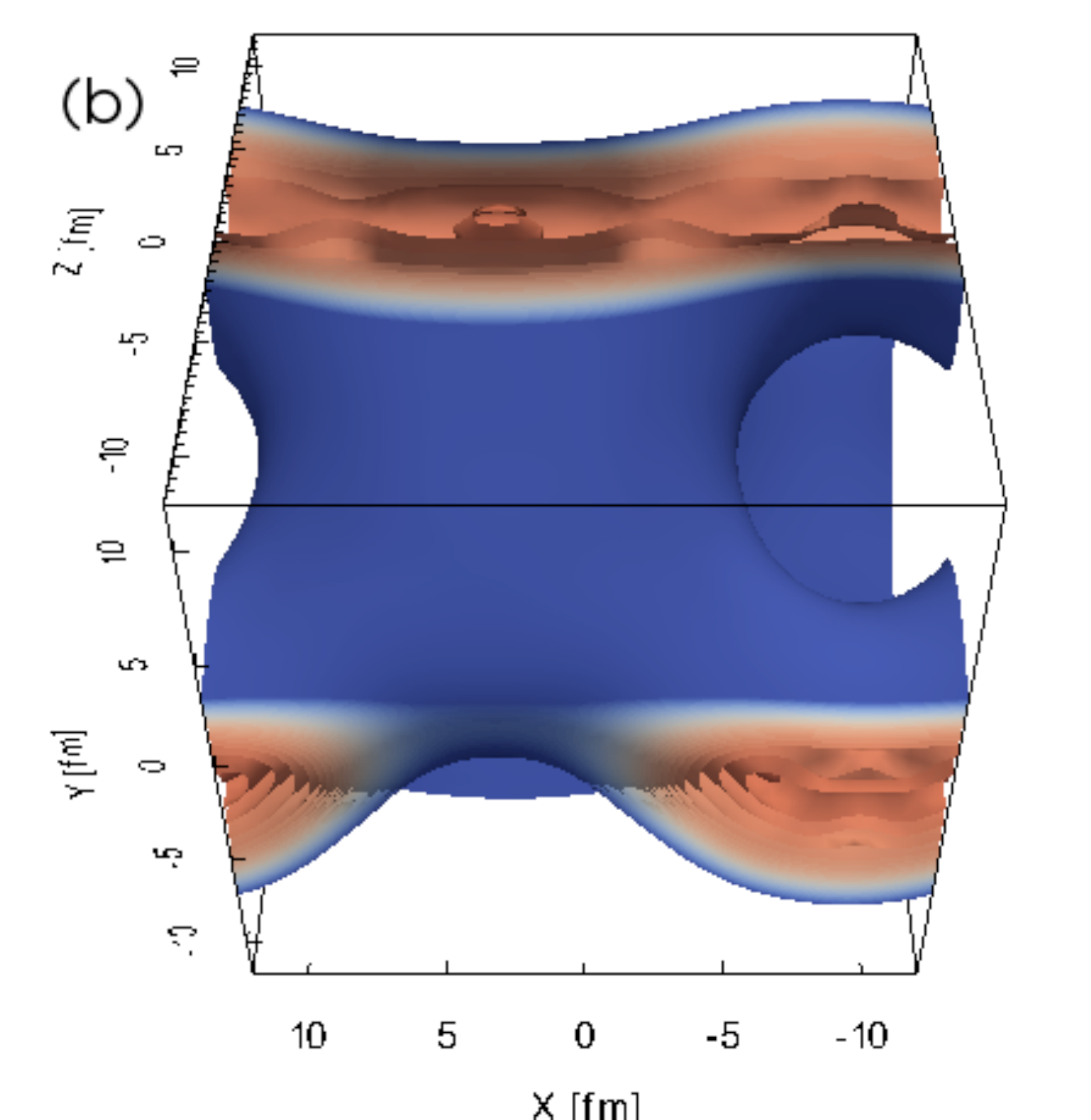}
\end{minipage}
\begin{minipage}{0.32\textwidth}
\includegraphics[width = 0.89\textwidth]{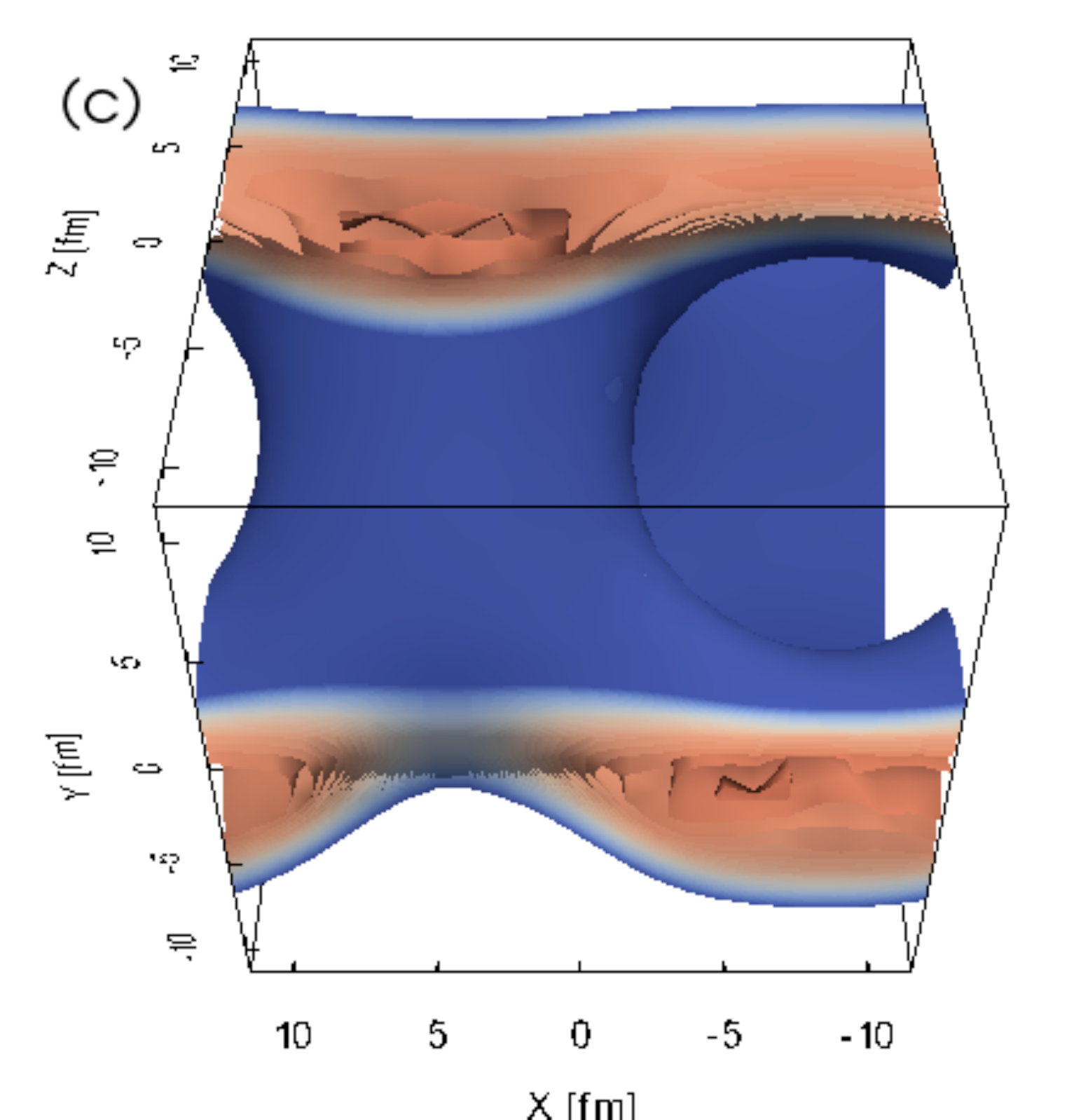}
\end{minipage}}
\caption{Iso-surfaces of $\rho (\vec{r})$ as in Fig.\ref{fig:pasta}. The orientation corresponds to the top plate of the converged waffle phase. Subfigure (a) shows the result of the M-SHF simulation without spin-orbit and (b) with spin-orbit interactions. Subfigure (c) is the converged Sky3D simulation with spin-orbit.}
\label{fig:pasta_so1}
\end{figure*}
\begin{figure*}
\centering{
\begin{minipage}{0.32\textwidth}
\includegraphics[width = 0.92\textwidth]{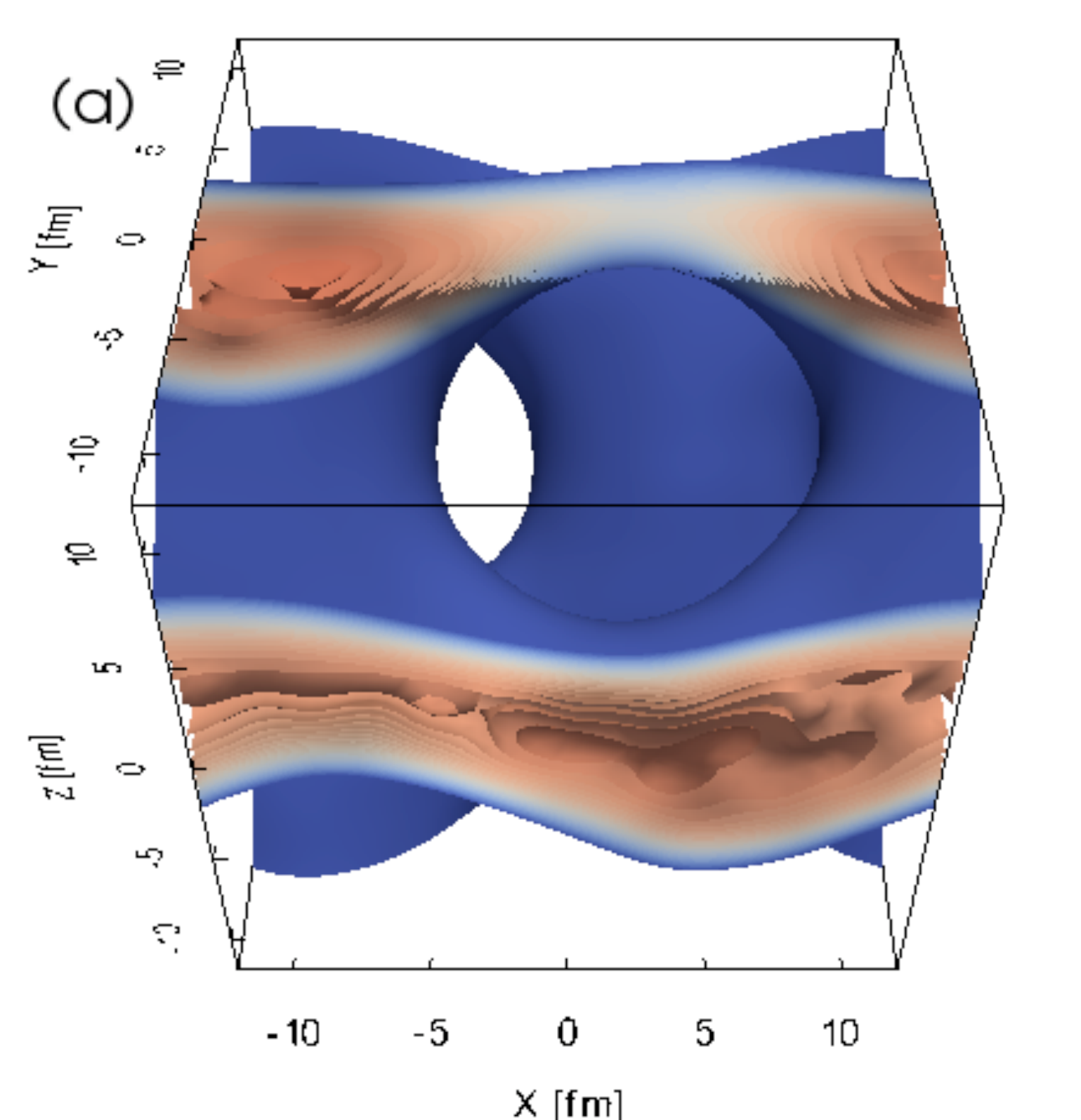}
\end{minipage}
\begin{minipage}{0.32\textwidth}
\includegraphics[width = 0.92\textwidth]{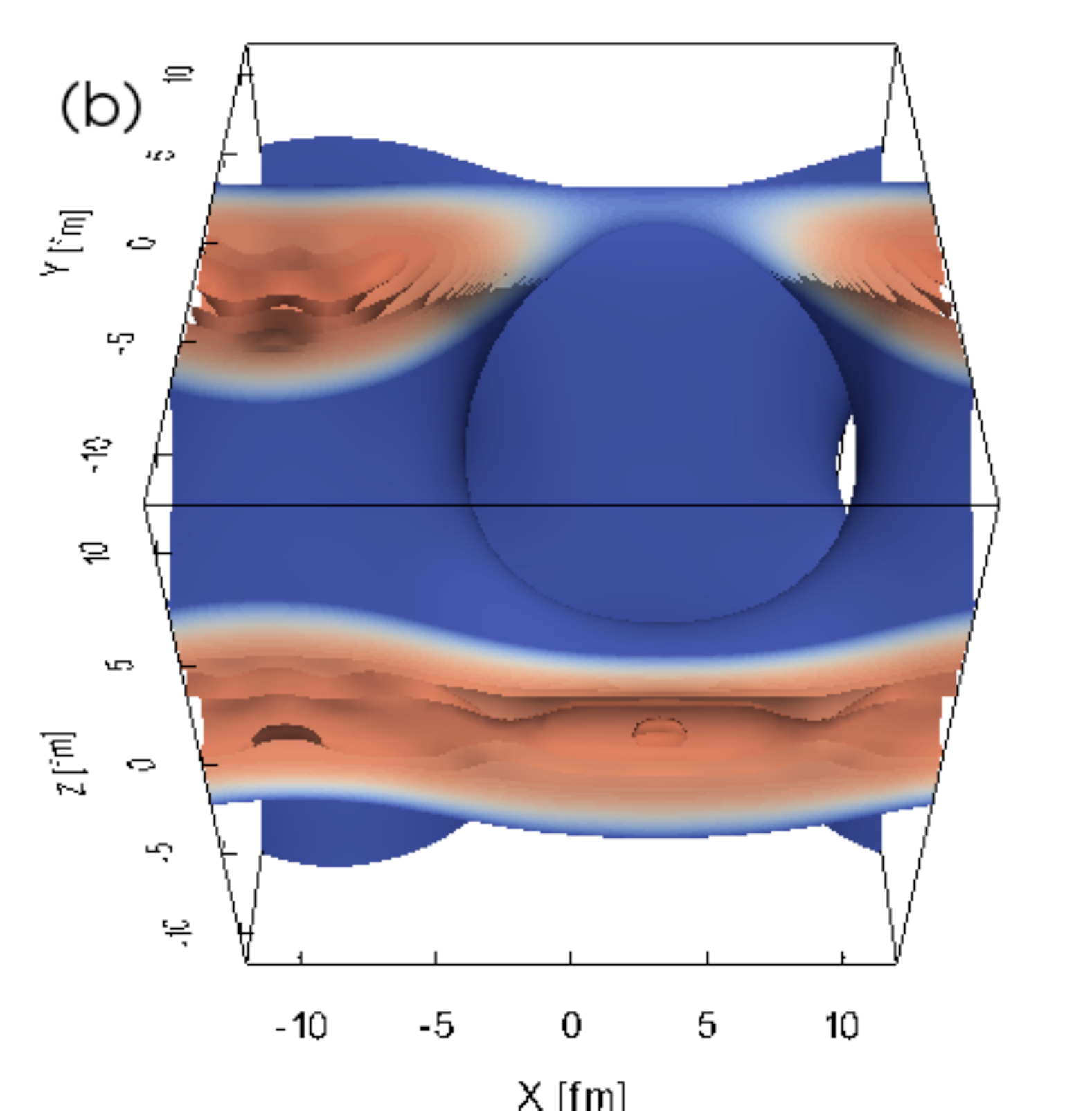}
\end{minipage}
\begin{minipage}{0.32\textwidth}
\includegraphics[width = 0.89\textwidth]{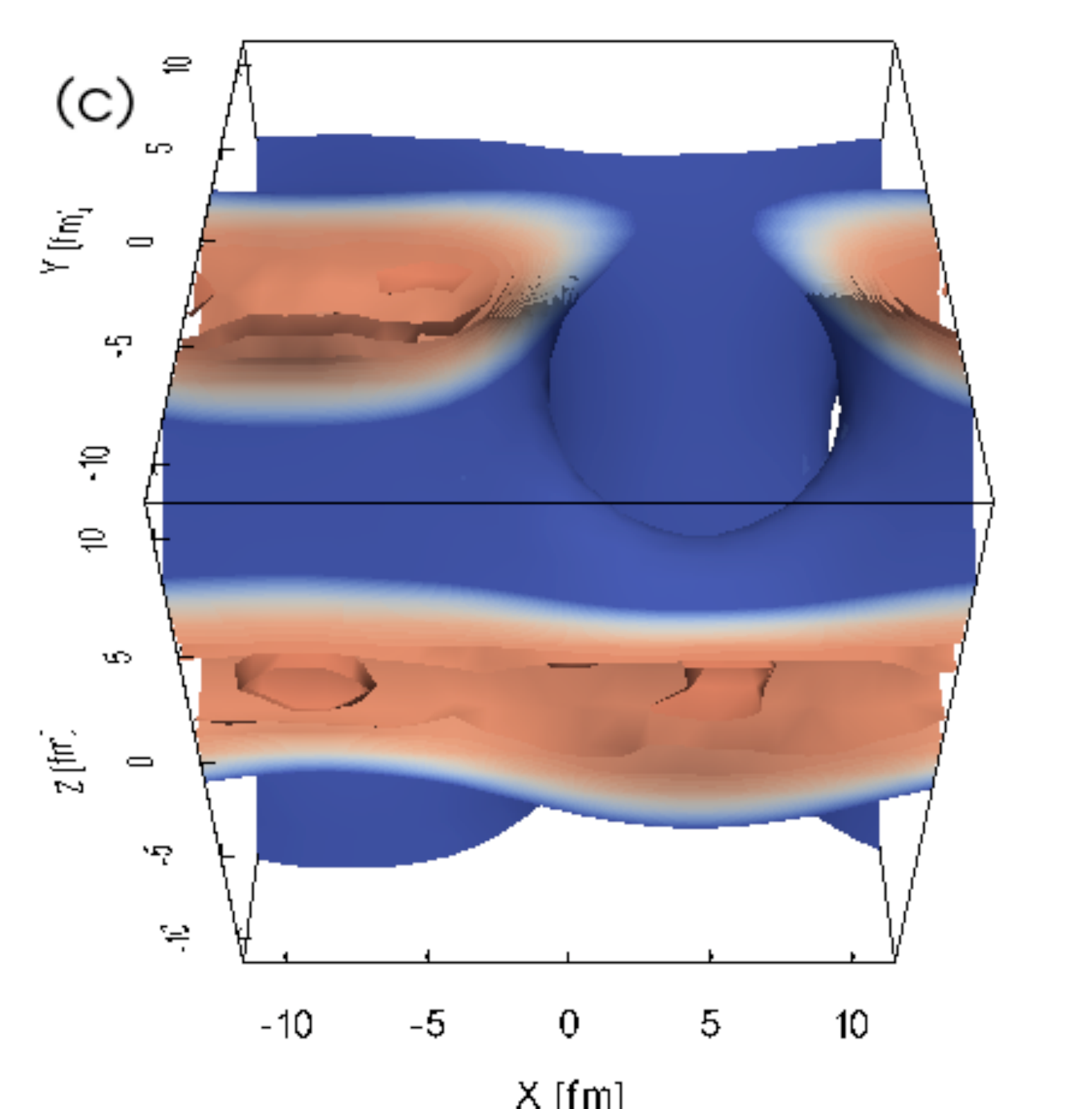}
\end{minipage}}
\caption{Iso-surfaces of $\rho (\vec{r})$ as in Fig.\ref{fig:pasta}. The orientation corresponds to the bottom plate of the converged waffle phase. Subfigure (a) shows the result of the M-SHF simulation without spin-orbit and (b) with spin-orbit interactions. Subfigure (c) is the converged Sky3D simulation with spin-orbit.}
\label{fig:pasta_so2}
\end{figure*}
\newline
Figures \ref{fig:pasta_so1} and \ref{fig:pasta_so2} show the converged pasta for M-SHF without and with spin-orbit terms and the Sky3D simulation with spin-orbit. Although the general shape is the same, we can find subtle differences. Without spin-orbit, the size of the holes in the top and bottom plates seem very similar. When the spin-orbit contributions are added, one hole shrinks while the other one becomes larger. This evolution corresponds to the shape changes that were indicated in Fig.\:\ref{MD_log_so} between iteration 10000 - 18000. Interestingly, the M-SHF and Sky3D simulations both show this effect despite the different initializations. For M-SHF, the small hole is in the top plate while the large one is in the bottom plate. For Sky3D the situation is reversed. However, due to the periodic boundary conditions the order is not important and the pasta phase should consist of a lattice of alternating small and large holes. The question is of course, whether the same structure would be found in simulations of a larger volumes or with different spin-orbit potentials. More systematic studies have to be performed in the future. At present, we conclude that the inclusion of the spin-orbit contribution in the Sv-bas nuclear potential modifies features of the waffle phase but does not lead to its disappearance. 
\section{Summary}
\label{summary}
In this work, we introduce and discuss calculations of nuclear matter via Skyrme Hartree-Fock calculations with the Multi-resolution ADaptive Numerical Environment for Scientific Simulations (MADNESS). To verify and benchmark the code, we perform calculations of nuclear ground states and find good agreement with the established Skyrme Hartree-Fock code Sky3D and experimental binding energies. While calculations for light nuclei seem to be very fast, the scaling of the code with number of nucleons needs improvement for future studies. We test our code for large boxes and free boundary conditions and small boxes with periodic boundary conditions. For nuclear pasta simulations, we explore a configuration in a 24\:fm box with periodic boundary conditions and 700 nucleons with a proton fraction of 0.3 and an average density of $\rho = 0.05\:\mathrm{fm}^{-3}$. The initialization of the simulation is done using the output of a converged simulation by the Indiana University Molecular Dynamics code. The corresponding shape is the waffle phase. For a calculation without spin-orbit terms, we find that the simulation fulfills our convergence criterium after 8000 iterations whereas the configuration and binding energies do not change significantly after iteration $\sim 4000$. Furthermore, the final shape of the nuclear configuration does not differ significantly from the initial MD state indicating that the waffle phase is stable even when quantum mechanical effects are considered. When adding spin-orbit nuclear potential terms to a partially converged calculation, we find small shape changes which push the convergence to iteration $\sim 20000$. The shape of the waffle phase has small but visible differences in comparison to the calculation without spin-orbit. However, the phase remains in the waffle geometry. Similar results are also found with the Sky3D calculation.  
\acknowledgments
The authors would like to thank M. Caplan for providing data from simulations with the Indiana University Molecular Dynamics (IUMD) code and Bastian Schuetrumpf for his assistance with the Molecular Dynamics simulation initialization for the Sky3d code. This work was supported in part by the Lilly Endowment, Inc., through its support for the Indiana University Pervasive Technology Institute, and in part by the Indiana METACyt Initiative. The Indiana METACyt Initiative at IU is also supported in part by the Lilly Endowment, Inc. This research used resources of the Oak Ridge Leadership Computing Facility at ORNL, which is supported by the Office of Science of the U.S. Department of Energy under Contract No. DE-AC05-00OR22725. This work was also supported by DOE grants DE-FG02-87ER40365 (Indiana University) and DE-SC0008808 (NUCLEI SciDAC Collaboration)
\bibliography{ref}
\end{document}